\documentclass[reprint,amsmath,amssymb,aps,floatfix]{revtex4-2}
\usepackage{graphicx}
\usepackage{dcolumn}
\usepackage{bm}
\usepackage[english]{babel}
\usepackage{float}

\begin{document}

\title{Stimulated Raman adiabatic passage in optomechanics}
 
\author{Vitaly Fedoseev$^{1}$, Fernando Luna$^2$, Ian Hedgepeth$^2$, Wolfgang L\"{o}ffler$^1$ and Dirk Bouwmeester$^{1,2}$}
 
\affiliation{%
 $^1$Huygens-Kamerlingh Onnes Laboratorium, Leiden University, 2333 CA Leiden, The Netherlands.\\
 $^2$Department of Physics, University of California, Santa Barbara, CA 93106, USA.
}%








\begin{abstract}

In multimode optomechanical systems, the mechanical modes can be coupled via the radiation pressure of the common optical mode, but the fidelity of the state transfer is limited by the optical cavity decay. Here we demonstrate stimulated Raman adiabatic passage (STIRAP) in optomechanics, where the optical mode is not populated during the coherent state transfer between the mechanical modes avoiding this decay channel. We show a state transfer of a coherent mechanical excitation between vibrational modes of a membrane in a high-finesse optical cavity with a transfer efficiency of 86\%. Combined with exceptionally high mechanical quality factors, STIRAP between mechanical modes can enable generation, storage and manipulation of long-lived mechanical quantum states, which is important for quantum information science and for the investigation of macroscopic quantum superpositions.

\end{abstract}

\maketitle

STIRAP describes adiabatic population transfer between two states coherently coupled via a mediating state that remains unoccupied. This renders STIRAP robust against loss and noise in the mediating state, leading to profound applications in atomic- and molecular-optics research \cite{Gaubatz1988, Pillet1993}, trapped-ion physics \cite{Sorensen2006}, superconducting circuits \cite{Kumar2016}, other solid-state systems \cite{Goto2007, Golter2014}, optics \cite{Longhi2009}, in entanglement generation \cite{Simon2007, Chang2020} and qubit operations \cite{Toyoda2013}. STIRAP in optomechanics has been proposed for optical frequency conversion with a mechanical mode being the mediating state, where the fidelity of the state transfer is not deteriorated by the residual thermal noise of the mechanical mode \cite{Wang2012a,Tian2012} and for a mechanical state transfer through the common optical mode \cite{Garg2017}.

State transfer between nondegenerate mechanical modes was demonstrated in \cite{Weaver2017, Xu2019} where the beating between two driving light fields bridges the frequency difference of the modes. The motion of the modes modulate the intracavity light fields creating motional sidebands \cite{Aspelmeyer2014}. This transfer scheme relies on the matched motional sidebands and requires the detuning of the driving fields to be much higher than the mechanical frequencies \cite{Buchmann2015}. In this case the other unmatched motional sidebands are of similar amplitudes as the matched ones and cause incoherent driving or cooling of the mechanical modes, limiting the state transfer fidelity. In optomechanical STIRAP in the sideband-resolved regime the loss due to the unmatched motional sidebands can be made negligibly small by choosing the detuning of the two driving light fields equal to the frequencies of the mechanical modes. In this case the two matched sidebands at the cavity resonance interfere destructively driving the state transfer and the other motional sidebands have much smaller amplitudes. This strongly decreases the unwanted effects of the unmatched motional sidebands and allows the state transfer fidelity to approach unity in the quantum regime.

Fig. 1(a) shows the basic $\Lambda$-type arrangement of 3 levels typical for STIRAP. In the triply rotating frame at frequencies $\omega_i=E_i/\hbar$ for states $\psi_i$, $i=1,2,3$, the Hamiltonian is: 
\begin{equation} \label{eq:1}
\hat{H}(t)=\frac{\hbar}{2}
\begin{pmatrix}
0 & \Omega_{12}(t) & 0 \\
\Omega_{12}(t) & 0 & \Omega_{23}(t) \\
0 & \Omega_{23}(t) & 0
\end{pmatrix},
\end{equation}
with $\Omega_{12}$ and $\Omega_{23}$ the Rabi frequencies resulting from two driving fields at frequencies $(E_2-E_1)/\hbar$ and $(E_2-E_3)/\hbar$. One of the three instantaneous eigenstates has eigenvalue 0 and does only include states $\psi_1$ and $\psi_3$:
\begin{equation}
\Phi_0(t)=\cos{\theta(t)}\psi_1-\sin{\theta(t)}\psi_3,
\end{equation}
with $\tan{\theta(t)}=\Omega_{12}(t)/\Omega_{23}(t)$. The existence of this "dark" state in optomechanics has been firstly demonstrated in \cite{Dong2012}. STIRAP is based on the adiabatic following of $\Phi_0(t)$ by slowly varying $\theta(t)$ from $\theta(-\infty)=0$ to $\theta(\infty)=\pi/2$. Thus, the system can be adiabatically transferred from $\psi_1$ to $\psi_3$ never occupying state $\psi_2$. Fig. 1(b) shows a driving pulse sequence satisfying this requirement and Fig. 1(c) shows the energy eigenvalues corresponding to the three eigenstates $\Phi_+(t)$, $\Phi_0(t)$, and $\Phi_-(t)$. This driving pulse sequence together with the adiabaticity condition $\sqrt{\Omega_{12}(t)^2+\Omega_{23}(t)^2}\gg \dot{\theta}$ prevents the lossy mediating state from being occupied throughout the transfer process.

The Hamiltonian in Eq. (\ref{eq:1}) can be realized in multimode optomechanics \cite{Wang2012a,Tian2012} where states 1 and 3 are mechanical excitations with frequencies $\omega_1$ and $\omega_2$ and state 2 is an optical cavity mode at $\omega_{\mathrm{cav}}$, see Fig. 1(d). Two optical driving fields at $\omega_{\mathrm{L}i}=\omega_{\mathrm{cav}}-\omega_i$ for $i=1,2$ are introduced in order to create the beamsplitter interaction $\hat{a}\hat{b}^\dagger_i + \hat{a}^\dagger\hat{b}_i$ that couples the mechanical modes to the cavity mode, where $\hat{a}$($\hat{a}^\dagger$) and $\hat{b_i}$($\hat{b}^\dagger_i$) are the photon and phonon annihilation (creation) operators.
\begin{figure}[ht]
\centering
\includegraphics[scale=1]{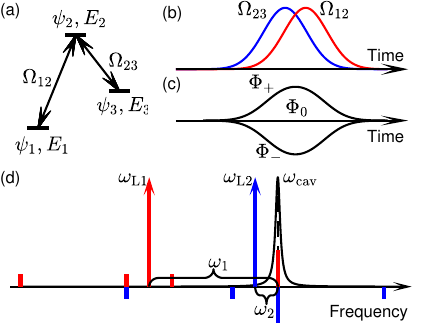}
\caption{STIRAP scheme in optomechanics. (a) Energy levels diagram. (b) Coupling strengths of the pulse sequence for the driving fields. (c) The resulting energy eigenvalues for the instantaneous Hamiltonian eigenstates. STIRAP explores the properties of $\Phi_0(t)$ given in Eq. (2). (d) The optomechanical implementation contains a cavity mode at frequency $\omega_{\mathrm{cav}}$, two driving fields at $\omega_{\mathrm{L}1}$ and $\omega_{\mathrm{L}2}$ and eight motional sidebands due to the mechanical modes at $\omega_1$ and $\omega_2$ on the driving fields, red bars corresponding to the sidebands on $\omega_{\mathrm{L}1}$ and blue bars corresponding to the sidebands on $\omega_{\mathrm{L}2}$. Two sidebands match $\omega_{\mathrm{cav}}$. In the case of $\Phi_0(t)$ the states $\psi_1$ and $\psi_3$ are out of phase leading to destructive interference of the sidebands that overlap with $\omega_{\mathrm{cav}}$.}
\end{figure}
\begin{figure}[ht]
\centering
\includegraphics[scale=1]{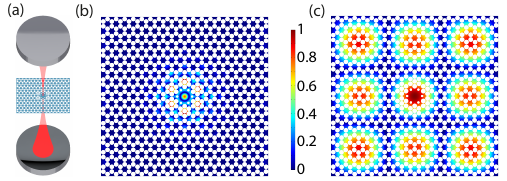}
\caption{Optomechanical setup. (a) A transparent dielectric membrane patterned with a phononic crystal is placed in the middle of a high-finesse optical cavity. Shift of the membrane along the axis of the cavity changes the cavity resonance frequency, causing coupling of light in the cavity to vibrational modes of the membrane. (b) Simulated displacement of a mechanical mode of the defect of the phononic crystal. The mode is localized as its frequency lies in the band gap (mode 1, initially excited). (c) Simulated displacement of the 3,3 drumhead mode of the full membrane (mode 2). This second mode was selected because it has an appropriate mechanical frequency and quality factor and has a maximum amplitude at the center. This allows both modes to be aligned for optimal coupling to the same cavity mode.}
\end{figure}
The optical mode can be represented by the operator $\hat{a}=\bar{\alpha}+\delta \hat{a}$, where $\bar{\alpha}$ is the average coherent amplitude due to the driving optical fields and $\delta \hat{a}$ is the fluctuating term \cite{Aspelmeyer2014}. Each mechanical mode produces two sidebands on each optical field. Due to resonance with the cavity the two sidebands with frequencies $\omega_{\mathrm{cav}}$ acquire much larger amplitudes than the other sidebands. Taking into account only those two sidebands and including mechanical and optical loss rates, $\Gamma_i$ and $\kappa$, the time evolution of the state vector $\psi(t)=(\hat{b}_1(t), \delta\hat{a}(t), \hat{b}_2(t))^T$ is given by
\begin{equation}
i\frac{d\psi(t)}{dt}=
\begin{pmatrix}
-i\frac{\Gamma_1}{2} & g_1(t) & 0 \\
g_1(t) & -i\frac{\kappa}{2} & g_2(t)\\
0 & g_2(t) & -i\frac{\Gamma_2}{2}
\end{pmatrix}\psi(t).
\end{equation}
Here, the rotating wave approximation has been used and it is valid in the linearized regime of cavity optomechanics \cite{Aspelmeyer2014}. $g_i(t)$ is the optomechanical multiphoton coupling for mechanical modes $i=1,2$, $g_i=g_{i0}\bar{\alpha}_i$, where $g_{i0}$ is single photon coupling and $\bar{\alpha}_i$ is the driving field at $\omega_{\mathrm{L}i}$, see Supplemental Material. Equation (3) is valid in the sideband resolved regime together with the requirement $|\omega_1-\omega_2|\gg \kappa$ and is identical to Eq. (1) in the absence of losses and with the Rabi frequencies $\Omega_{12}$ and $\Omega_{23}$ corresponding to $2g_1$ and $2g_2$. 

Experimentally we demonstrate the state transfer in the membrane-in-the-middle (MIM) configuration \cite{Thompson2008}, where a membrane with low optical absorption is placed in the center of a high-finesse optical cavity with $\kappa/2\pi=54$ $\mathrm{kHz}$ (including membrane), see Fig. 2. A displacement of the membrane along the optical axis leads to a shift in the optical cavity transmission described by the interaction Hamiltonian $\hat{H}_{int}=-\hbar g_0 \hat{a}^\dagger \hat{a}(\hat{b}+\hat{b}^\dagger)$, where $g_0$ is the single photon optomechanical coupling \cite{Aspelmeyer2014}. The membrane is a highly stressed 25 nm thick SiN film lithographically patterned with a phononic crystal with a defect in its center suspended on a silicon frame \cite{Tsaturyan2017}. There are two types of mechanical modes: whole membrane drumhead modes and modes localized near the phononic crystal defect with frequencies in the phononic crystal bandgap. Vibrational energy of the drumhead modes is mainly lost in the bending regions where the membrane is connected to the frame \cite{Schmid2011, Tsaturyan2017}. The modes localized near the defect possess enhanced quality factors by 1-2 orders of magnitude compared to the drumhead modes \cite{Tsaturyan2017}. We demonstrate STIRAP between the fundamental mode of the defect with frequency $\omega_1/2\pi=1.25$ $\mathrm{MHz}$ and quality factor $Q=1.3\cdot 10^7$ (mode 1, Fig. 2(b)), and the 3,3 drumhead mode with $\omega_2/2\pi=0.22$ $\mathrm{MHz}$ and $Q=1.2\cdot 10^6$ (mode 2, Fig. 2(c)). The modes are coupled to the optical cavity with single-photon couplings of $g_{01}/2\pi=1.5\pm 0.1$ $\mathrm{Hz}$ and $g_{02}/2\pi=1.0\pm 0.1$ $\mathrm{Hz}$ respectively. In addition to these modes possessing relatively large single photon coupling, quality factors and frequency separation, there are no other mechanical modes in the range of $\sim 1/\sigma$ from $\omega_1$ and $\omega_2$, where $\sigma$ defines the width of the driving pulses. The latter requirement guaranties that modes 1 and 2 are not coupled to other modes during the transfer. STIRAP is very sensitive to the double-photon detuning $\Delta_{2\mathrm{ph}}=(\omega_{\mathrm{L}1}+\omega_1)-(\omega_{\mathrm{L}2}+\omega_2)$ \cite{Vitanov2017}, therefore the two optical fields are created by amplitude modulation of light from a single 1064 nm laser using acousto-optic modulator (AOM). As a result the laser phase noise is not limiting the transfer efficiency \cite{Kuhn1992}. Due to nonlinear response of the AOM the detuning of this single laser light tone is chosen such that harmonics of the AC voltage sent to the AOM have a negligible effect on the transfer efficiency (see Supplemental Material). The membrane is in a vacuum chamber with pressure below $10^{-6}$ mbar at room temperature.
\begin{figure}[ht]
\centering
\includegraphics[scale=1]{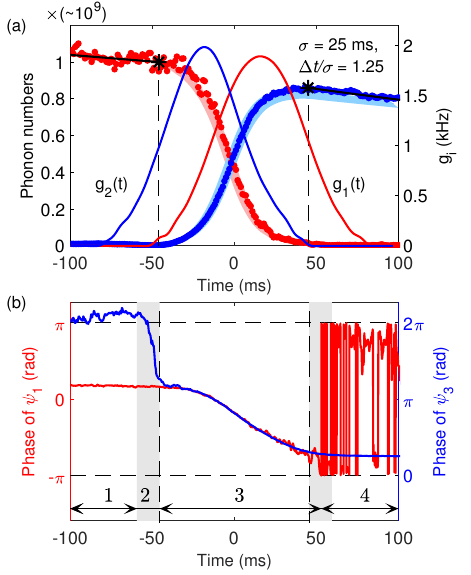}
\caption{Experimental optomechanical STIRAP. (a) Left scale: phonon number as a function of time, red dots correspond to averaged measurements for mode 1 ($\psi_1$), blue dots for mode 2 ($\psi_3$). The prefactor $10^9$ is a rough estimate. Light red and light blue regions represent the phonon populations with statistical uncertainties (1 standard deviation) obtained from simulations without free fit parameters. Right scale: multiphoton optomechanical coupling strengths, calculated from measured pulse intensities. The driving field pulses have a nearly Gaussian profile with the standard deviation parameter $\sigma$ and separation $\Delta t$, but their beginning and ending are smoothly truncated to zero. Black stars correspond to the phonon populations used to calculate transfer efficiency (5\% of the peak voltage sent to the AOM). (b) Measured phases of mode 1 (red) and mode 2 (blue) in the rotating frame.}
\label{fig:fig2}
\end{figure}
\begin{figure*}[ht]
\centering
\includegraphics[scale=1]{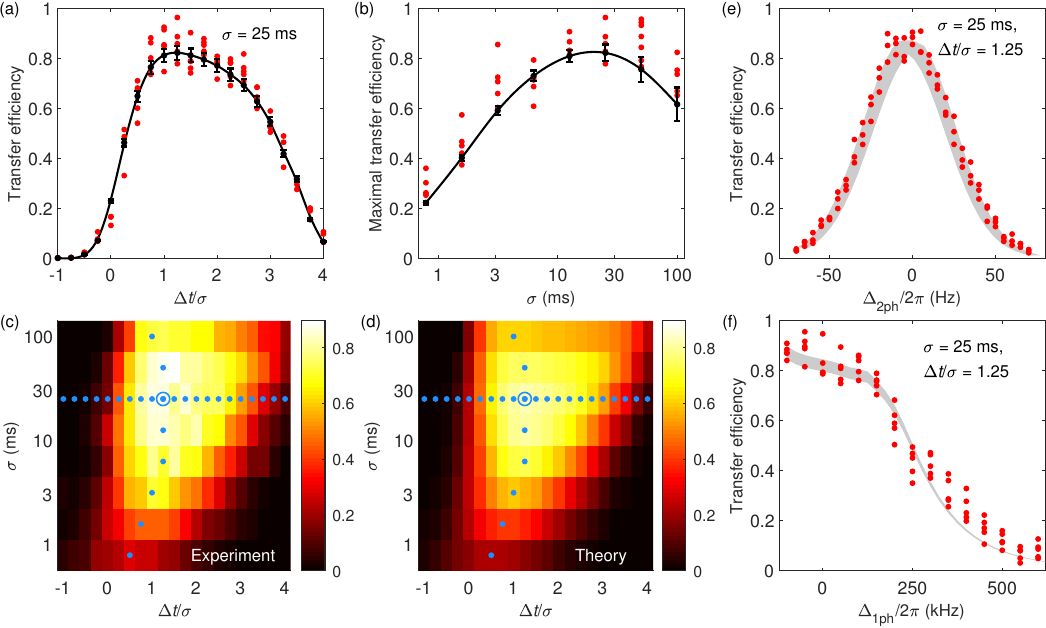}     
\caption{Transfer efficiencies under different parameters of the optical pulses. (a) The transfer efficiency as a function of the ratio of the delay between the pulses $\Delta t$ and the Gaussian pulse width $\sigma$. Positive values of $\Delta t$ correspond to the case that the field at $\omega_{\mathrm{L}2}$ is applied before the field at $\omega_{\mathrm{L}1}$. (b) Maximal transfer efficiencies as a function of $\sigma$. In (a) and (b) the red dots show measured efficiencies in individual runs, black dots are the simulated efficiencies, and the black lines are guides to the eye. The increasing error bars for larger $\sigma$ in (b) are due to observed but not accounted for small non-linear and heating effects, see Supplemental Material. (c) and (d) show the experimental (c) and predicted (d) transfer efficiency as a function of the Gaussian parameter $\sigma$ and separation $\Delta t$. The horizontal row of dots in (c) and (d) correspond to the data shown in (a), while the vertical row of dots correspond to the data shown in (b). The transfer process for the parameters corresponding to the open circle in (c) and (d) is shown in Fig. 3(e). The efficiency as a function of the two-photon detuning $\Delta_{2\mathrm{ph}}$ with zero single-photon detuning. (f) The efficiency as a function of the single-photon detuning $\Delta_{1\mathrm{ph}}$ with zero two-photon detuning. In (e) and (f) the red circles are measured efficiencies in individual runs, and the shaded regions are simulated efficiencies with statistical uncertainties. The simulated curve in (e) has a frequency correction of 4 Hz caused by small non-linear and heating effects, see Supplemental Material.}
\label{fig:fig4}
\end{figure*}

STIRAP with parameters tuned for maximum phonon number state transfer efficiency is shown in Fig. \ref{fig:fig2}. The measurement of a typical transfer process has the following sequence: mode 1 is excited to an amplitude much higher than its thermal occupation by applying an AC voltage in resonance with the mechanical frequency to a needle positioned close to the center of the membrane. During its free decay the two optical pulses are sent which transfers the excitation of mode 1 to mode 2. The transfer starts with the beginning of pulse 1 (red) and finishes with the end of pulse 2 (blue), these moments are denoted by dashed vertical lines. The transfer efficiency is calculated as the ratio of the number of phonons in mechanical mode 2 at the end of the transfer to the number of phonons in mechanical mode 1 at the beginning of the transfer (black stars). A theoretical model without free fit parameters was developed in the classical limit to simulate the transfer process taking into account the corrections due to the other sidebands and the measured profiles of the light pulses (see Supplemental Material), and shows excellent agreement to the experimental data in Fig. 3(a). Simulations show that the average rate of loss through the optical mode is $\sim $1 Hz in the dark state during the transfer. We observe small variations in the frequencies of the mechanical modes with each STIRAP sequence. To account for these variations, we measure the mechanical frequencies in thermal motion and adjust the values of the mechanical frequencies for the driving pulse generation accordingly before each STIRAP sequence.

In our realization of STIRAP using coherent state populations, i.e. in the classical regime, the phases of the mechanical modes during the transfer can be continuously monitored, see Fig. 3(b). There are four time domains with distinct behavior of phases: in domain 1 $g_1(t)=0$ and the phase of mode 1 is defined by the excitation used to drive it, while mode 2 is in its thermal motion, thus the difference between the phases is random; in domain 2 STIRAP starts and the phase of mode 2 adjusts itself until the sidebands at $\omega_{\mathrm{cav}}$ become $\pi$ out of phase; in domain 3 the phase of the locked mechanical modes changes due to the optomechanically induced frequency shift from field $\omega_{\mathrm{L}2}$ (unmatched sidebands); in domain 4 the read-out signal of mode 1 becomes much less than the read-out noise.

Next we investigate the dependence of the transfer efficiency on the parameters of the process. First the time delay between the optical pulses $\Delta t$ is varied, see Fig. 4. The adiabaticity condition becomes more and more violated when the separation between the pulses is too small or too large, leading to decreasing efficiency. Then the duration of the pulses $\sigma$ is varied while keeping the time delay $\Delta t$ optimal. The adiabaticity condition is satisfied increasingly better with longer pulses such that for pulses with $\sigma=100$ $\mathrm{ms}$ only 2\% of the initial phonon population in mode 1 is lost through the population and decay of the optical mode. Nevertheless the efficiency starts to decrease for $\sigma\gtrsim{25}$ $\mathrm{ms}$ due to the mechanical decay of the modes, setting the upper bound on the transfer efficiency. The solid curves in Fig. 4(a-b) are numerical results and Fig. 4(c-d) compare experiment and simulations for varying $\Delta t$ and $\sigma$. We observe an increasing discrepancy between measured and simulated data for the state transfer with $\sigma\gtrsim{25}$ ms. This is caused by membrane heating from the driving pulses and by the defect mode frequency dependence on the amplitude of the full membrane 3,3 mode, see Supplemental Material.

A signature of STIRAP \cite{Vitanov2017} is strong sensitivity of the transfer efficiency to the two-photon detuning $\Delta_{2\mathrm{ph}}=(\omega_{\mathrm{L}1}+\omega_1)-(\omega_{\mathrm{L}2}+\omega_2)$ given $\Delta_{1\mathrm{ph}}=0$, compared to the sensitivity to the single-photon detuning $\Delta_{1\mathrm{ph}}=\omega_{\mathrm{cav}}-(\omega_{\mathrm{L}1}+\omega_1)$ given $\Delta_{2\mathrm{ph}}=0$, Fig. 4(e-f). The frequency scale for the two-photon detuning is set by the duration of the transfer process: $\Delta_{2\mathrm{ph}} \sim \pi/T_{\mathrm{transfer}}$, implying that the sidebands at $\omega_{\mathrm{cav}}$ accumulate a phase difference of $\pi$ during the transfer and consequently no longer interfere destructively. The frequency scale for $\Delta_{1\mathrm{ph}}$ is set by the optical cavity linewidth $\kappa$: non-zero $\Delta_{1\mathrm{ph}}$ leads to changes in the intracavity light fields intensities and in the amplitudes of the sidebands.


The highest phonon number transfer efficiency we observe in our system is $86\pm3$\%. The highest demonstrated state transfer efficiencies in other systems are: transmon qubit 83\% \cite{Kumar2016}; BEC of atoms 87\% \cite{Dupont-Nivet2015}; trapped ions 90\% \cite{Sorensen2006}; superconducting Xmon qutrit 96\% \cite{Xu2016a}; doped crystals $98\pm2$\% \cite{Klein2008}; atom beams $98\pm2$\% \cite{Theuer1998}. In general, the STIRAP scheme in optomechanics can result in transfer efficiencies close to unity provided that the difference between the frequencies of the mechanical modes is much larger than the cavity linewidth, while being in the weak coupling regime, with peak multiphoton optomechanical couplings being much larger than the inverse of the transfer duration, and with slow enough mechanical decay. This allows us to formulate the following requirements for optomechanical STIRAP:
\begin{equation}
|\omega_1-\omega_2| \gg \kappa \gg \max{g_i(t)} \gg 2\pi/T_{\mathrm{transfer}} \gg \Gamma_i.
\end{equation}
This set of stringent requirements applies both to the classical and the quantum regime of STIRAP in optomechanics. Other experimental challenges are the accurate control of 1- and 2-photon detunings, circumventing detrimental effects of the unmatched sidebands, and proving stable subwavelength positioning of the membrane to maximize the coupling strength. 


Further we show that state transfer via STIRAP of 1-phonon Fock state can in principle be observed experimentally with the same membranes in a cryogenic setting, for details see Supplemental Material. We provide a full quantum treatment of the protocol for such a state transfer including known sources of noise and unwanted effects: thermalization to the environment, heating by the laser light fields, presence of other nearby membrane modes, realistic overall detection efficiency and dark count rate of a single photon detector. We consider STIRAP between two modes of the defect of the phononic crystal in the membrane, with quality factors of $10^9$ \cite{Rossi2018}, resulting in a thermal decoherence time \cite{Aspelmeyer2014} of approximately 5 ms at 1 K. In laser cooling experiments \cite{Peterson2016,Underwood2015} the steady state temperature of similar membranes was observed to be less than 0.5 K above the cryostat base temperature when being sideband cooled, thus we adopt 1K as a conservative  estimate for the membrane temperature due to laser heating. 

The protocol consists of the following steps: both modes are sideband cooled to an average phonon occupation $\bar{n}=0.1$; detection of a Stokes photon from a short blue-detuned pulse projects the state of mode 1 to a state close to 1-phonon Fock state; the STIRAP pulse sequence is sent; the state of the modes is read-out by a short red-detuned pulse through detection of anti-Stokes photons. It is essential to filter out the strong pump light fields and to send the scattered photons to a single photon detector with high enough overall detection efficiency. Based on demonstrated experimental parameters we calculate that a 1-phonon Fock state can be transferred with fidelity of 60\%.

In conclusion, in this Letter we have shown the first optomechanical implementation of STIRAP and demonstrated a maximum phonon number state transfer efficiency of $86\pm3$\%. The efficiency is benchmarked against variation in the STIRAP pulse duration and separation as well as against the STIRAP single- and 2-photon detuning and is found to be in good agreement with theory. Our quantum simulations show that STIRAP of a single phonon Fock state is feasible to observe with demonstrated technology. Furthermore, modified versions of STIRAP (fractional STIRAP \cite{Marte1991}, tripod STIRAP \cite{Unanyan1998}) can be used to create and detect entangled mechanical states. Therefore, STIRAP in optomehanics can play an important role in quantum information protocols and in generating macroscopic superposition states. 
\\
\\
We would like to thank K. Heeck, H. van der Meer and S. Sonar for support of this project. This work was supported by the UC Santa Barbara NSF Quantum Foundry funded via the Q-AMASE-i program Grant No. DMR-1906325, and the NWO Gravitation-grant Quantum Software Consortium - 024.003.037, and the NWO Vrij Programma Grant QUAKE - 680.92.18.04.
\bibliography{stirap.bib}

\begin{thebibliography}{32}%
\makeatletter
\providecommand \@ifxundefined [1]{%
 \@ifx{#1\undefined}
}%
\providecommand \@ifnum [1]{%
 \ifnum #1\expandafter \@firstoftwo
 \else \expandafter \@secondoftwo
 \fi
}%
\providecommand \@ifx [1]{%
 \ifx #1\expandafter \@firstoftwo
 \else \expandafter \@secondoftwo
 \fi
}%
\providecommand \natexlab [1]{#1}%
\providecommand \enquote  [1]{``#1''}%
\providecommand \bibnamefont  [1]{#1}%
\providecommand \bibfnamefont [1]{#1}%
\providecommand \citenamefont [1]{#1}%
\providecommand \href@noop [0]{\@secondoftwo}%
\providecommand \href [0]{\begingroup \@sanitize@url \@href}%
\providecommand \@href[1]{\@@startlink{#1}\@@href}%
\providecommand \@@href[1]{\endgroup#1\@@endlink}%
\providecommand \@sanitize@url [0]{\catcode `\\12\catcode `\$12\catcode
  `\&12\catcode `\#12\catcode `\^12\catcode `\_12\catcode `\%12\relax}%
\providecommand \@@startlink[1]{}%
\providecommand \@@endlink[0]{}%
\providecommand \url  [0]{\begingroup\@sanitize@url \@url }%
\providecommand \@url [1]{\endgroup\@href {#1}{\urlprefix }}%
\providecommand \urlprefix  [0]{URL }%
\providecommand \Eprint [0]{\href }%
\providecommand \doibase [0]{https://doi.org/}%
\providecommand \selectlanguage [0]{\@gobble}%
\providecommand \bibinfo  [0]{\@secondoftwo}%
\providecommand \bibfield  [0]{\@secondoftwo}%
\providecommand \translation [1]{[#1]}%
\providecommand \BibitemOpen [0]{}%
\providecommand \bibitemStop [0]{}%
\providecommand \bibitemNoStop [0]{.\EOS\space}%
\providecommand \EOS [0]{\spacefactor3000\relax}%
\providecommand \BibitemShut  [1]{\csname bibitem#1\endcsname}%
\let\auto@bib@innerbib\@empty
\bibitem [{\citenamefont {Gaubatz}\ \emph {et~al.}(1988)\citenamefont
  {Gaubatz}, \citenamefont {Rudecki}, \citenamefont {Becker}, \citenamefont
  {Schiemann}, \citenamefont {K{\"{u}}lz},\ and\ \citenamefont
  {Bergmann}}]{Gaubatz1988}%
  \BibitemOpen
  \bibfield  {author} {\bibinfo {author} {\bibfnamefont {U.}~\bibnamefont
  {Gaubatz}}, \bibinfo {author} {\bibfnamefont {P.}~\bibnamefont {Rudecki}},
  \bibinfo {author} {\bibfnamefont {M.}~\bibnamefont {Becker}}, \bibinfo
  {author} {\bibfnamefont {S.}~\bibnamefont {Schiemann}}, \bibinfo {author}
  {\bibfnamefont {M.}~\bibnamefont {K{\"{u}}lz}},\ and\ \bibinfo {author}
  {\bibfnamefont {K.}~\bibnamefont {Bergmann}},\ }\bibfield  {title} {\bibinfo
  {title} {{Population switching between vibrational levels in molecular
  beams}},\ }\href {https://doi.org/10.1016/0009-2614(88)80364-6} {\bibfield
  {journal} {\bibinfo  {journal} {Chemical Physics Letters}\ }\textbf {\bibinfo
  {volume} {149}},\ \bibinfo {pages} {463} (\bibinfo {year}
  {1988})}\BibitemShut {NoStop}%
\bibitem [{\citenamefont {Pillet}(1993)}]{Pillet1993}%
  \BibitemOpen
  \bibfield  {author} {\bibinfo {author} {\bibfnamefont {P.}~\bibnamefont
  {Pillet}},\ }\bibfield  {title} {\bibinfo {title} {{Adiabatic population
  transfer in a multilevel system}},\ }\href@noop {} {\bibfield  {journal}
  {\bibinfo  {journal} {Physical Review A - Atomic, Molecular, and Optical
  Physics}\ }\textbf {\bibinfo {volume} {48}},\ \bibinfo {pages} {1} (\bibinfo
  {year} {1993})}\BibitemShut {NoStop}%
\bibitem [{\citenamefont {S{\o}rensen}\ \emph {et~al.}(2006)\citenamefont
  {S{\o}rensen}, \citenamefont {M{\o}ller}, \citenamefont {Iversen},
  \citenamefont {Thomsen}, \citenamefont {Jensen}, \citenamefont {Staanum},
  \citenamefont {Voigt},\ and\ \citenamefont {Drewsen}}]{Sorensen2006}%
  \BibitemOpen
  \bibfield  {author} {\bibinfo {author} {\bibfnamefont {J.~L.}\ \bibnamefont
  {S{\o}rensen}}, \bibinfo {author} {\bibfnamefont {D.}~\bibnamefont
  {M{\o}ller}}, \bibinfo {author} {\bibfnamefont {T.}~\bibnamefont {Iversen}},
  \bibinfo {author} {\bibfnamefont {J.~B.}\ \bibnamefont {Thomsen}}, \bibinfo
  {author} {\bibfnamefont {F.}~\bibnamefont {Jensen}}, \bibinfo {author}
  {\bibfnamefont {P.}~\bibnamefont {Staanum}}, \bibinfo {author} {\bibfnamefont
  {D.}~\bibnamefont {Voigt}},\ and\ \bibinfo {author} {\bibfnamefont
  {M.}~\bibnamefont {Drewsen}},\ }\bibfield  {title} {\bibinfo {title}
  {{Efficient coherent internal state transfer in trapped ions using stimulated
  Raman adiabatic passage}},\ }\bibfield  {journal} {\bibinfo  {journal} {New
  Journal of Physics}\ }\textbf {\bibinfo {volume} {8}},\ \href
  {https://doi.org/10.1088/1367-2630/8/11/261} {10.1088/1367-2630/8/11/261}
  (\bibinfo {year} {2006})\BibitemShut {NoStop}%
\bibitem [{\citenamefont {Kumar}\ \emph {et~al.}(2016)\citenamefont {Kumar},
  \citenamefont {Veps{\"{a}}la{\"{i}}nen}, \citenamefont {Danilin},\ and\
  \citenamefont {Paraoanu}}]{Kumar2016}%
  \BibitemOpen
  \bibfield  {author} {\bibinfo {author} {\bibfnamefont {K.~S.}\ \bibnamefont
  {Kumar}}, \bibinfo {author} {\bibfnamefont {A.}~\bibnamefont
  {Veps{\"{a}}la{\"{i}}nen}}, \bibinfo {author} {\bibfnamefont
  {S.}~\bibnamefont {Danilin}},\ and\ \bibinfo {author} {\bibfnamefont {G.~S.}\
  \bibnamefont {Paraoanu}},\ }\bibfield  {title} {\bibinfo {title} {{Stimulated
  Raman adiabatic passage in a three-level superconducting circuit}},\ }\href
  {https://doi.org/10.1038/ncomms10628} {\bibfield  {journal} {\bibinfo
  {journal} {Nature Communications}\ }\textbf {\bibinfo {volume} {7}},\
  \bibinfo {pages} {1} (\bibinfo {year} {2016})}\BibitemShut {NoStop}%
\bibitem [{\citenamefont {Goto}\ and\ \citenamefont
  {Ichimura}(2007)}]{Goto2007}%
  \BibitemOpen
  \bibfield  {author} {\bibinfo {author} {\bibfnamefont {H.}~\bibnamefont
  {Goto}}\ and\ \bibinfo {author} {\bibfnamefont {K.}~\bibnamefont
  {Ichimura}},\ }\bibfield  {title} {\bibinfo {title} {{Observation of coherent
  population transfer in a four-level tripod system with a
  rare-earth-metal-ion-doped crystal}},\ }\href
  {https://doi.org/10.1103/PhysRevA.75.033404} {\bibfield  {journal} {\bibinfo
  {journal} {Physical Review A - Atomic, Molecular, and Optical Physics}\
  }\textbf {\bibinfo {volume} {75}},\ \bibinfo {pages} {1} (\bibinfo {year}
  {2007})}\BibitemShut {NoStop}%
\bibitem [{\citenamefont {Golter}\ and\ \citenamefont
  {Wang}(2014)}]{Golter2014}%
  \BibitemOpen
  \bibfield  {author} {\bibinfo {author} {\bibfnamefont {D.~A.}\ \bibnamefont
  {Golter}}\ and\ \bibinfo {author} {\bibfnamefont {H.}~\bibnamefont {Wang}},\
  }\bibfield  {title} {\bibinfo {title} {{Optically driven rabi oscillations
  and adiabatic passage of single electron spins in diamond}},\ }\href
  {https://doi.org/10.1103/PhysRevLett.112.116403} {\bibfield  {journal}
  {\bibinfo  {journal} {Physical Review Letters}\ }\textbf {\bibinfo {volume}
  {112}},\ \bibinfo {pages} {1} (\bibinfo {year} {2014})}\BibitemShut {NoStop}%
\bibitem [{\citenamefont {Longhi}(2009)}]{Longhi2009}%
  \BibitemOpen
  \bibfield  {author} {\bibinfo {author} {\bibfnamefont {S.}~\bibnamefont
  {Longhi}},\ }\bibfield  {title} {\bibinfo {title} {{Quantum-optical analogies
  using photonic structures}},\ }\href {https://doi.org/10.1002/lpor.200810055}
  {\bibfield  {journal} {\bibinfo  {journal} {Laser and Photonics Reviews}\
  }\textbf {\bibinfo {volume} {3}},\ \bibinfo {pages} {243} (\bibinfo {year}
  {2009})}\BibitemShut {NoStop}%
\bibitem [{\citenamefont {Simon}\ \emph {et~al.}(2007)\citenamefont {Simon},
  \citenamefont {Tanji}, \citenamefont {Ghosh},\ and\ \citenamefont
  {Vuletic}}]{Simon2007}%
  \BibitemOpen
  \bibfield  {author} {\bibinfo {author} {\bibfnamefont {J.}~\bibnamefont
  {Simon}}, \bibinfo {author} {\bibfnamefont {H.}~\bibnamefont {Tanji}},
  \bibinfo {author} {\bibfnamefont {S.}~\bibnamefont {Ghosh}},\ and\ \bibinfo
  {author} {\bibfnamefont {V.}~\bibnamefont {Vuletic}},\ }\bibfield  {title}
  {\bibinfo {title} {{Single-photon bus connecting spin-wave quantum
  memories}},\ }\href {https://doi.org/10.1038/nphys726} {\bibfield  {journal}
  {\bibinfo  {journal} {Nature Physics}\ }\textbf {\bibinfo {volume} {3}},\
  \bibinfo {pages} {765} (\bibinfo {year} {2007})}\BibitemShut {NoStop}%
\bibitem [{\citenamefont {Chang}\ \emph {et~al.}(2020)\citenamefont {Chang},
  \citenamefont {Zhong}, \citenamefont {Bienfait}, \citenamefont {Chou},
  \citenamefont {Conner}, \citenamefont {Dumur}, \citenamefont {Grebel},
  \citenamefont {Peairs}, \citenamefont {Povey}, \citenamefont {Satzinger},\
  and\ \citenamefont {Cleland}}]{Chang2020}%
  \BibitemOpen
  \bibfield  {author} {\bibinfo {author} {\bibfnamefont {H.-S.}\ \bibnamefont
  {Chang}}, \bibinfo {author} {\bibfnamefont {Y.}~\bibnamefont {Zhong}},
  \bibinfo {author} {\bibfnamefont {A.}~\bibnamefont {Bienfait}}, \bibinfo
  {author} {\bibfnamefont {M.-H.}\ \bibnamefont {Chou}}, \bibinfo {author}
  {\bibfnamefont {C.~R.}\ \bibnamefont {Conner}}, \bibinfo {author}
  {\bibfnamefont {{\'{E}}.}~\bibnamefont {Dumur}}, \bibinfo {author}
  {\bibfnamefont {J.}~\bibnamefont {Grebel}}, \bibinfo {author} {\bibfnamefont
  {G.~A.}\ \bibnamefont {Peairs}}, \bibinfo {author} {\bibfnamefont {R.~G.}\
  \bibnamefont {Povey}}, \bibinfo {author} {\bibfnamefont {K.~J.}\ \bibnamefont
  {Satzinger}},\ and\ \bibinfo {author} {\bibfnamefont {A.~N.}\ \bibnamefont
  {Cleland}},\ }\bibfield  {title} {\bibinfo {title} {{Remote entanglement via
  adiabatic passage using a tunably-dissipative quantum communication
  system}},\ }\href {https://doi.org/10.1103/PhysRevLett.124.240502} {\bibfield
   {journal} {\bibinfo  {journal} {Physical Review Letters}\ }\textbf {\bibinfo
  {volume} {124}},\ \bibinfo {pages} {240502} (\bibinfo {year} {2020})},\
  \Eprint {https://arxiv.org/abs/2005.12334} {arXiv:2005.12334} \BibitemShut
  {NoStop}%
\bibitem [{\citenamefont {Toyoda}\ \emph {et~al.}(2013)\citenamefont {Toyoda},
  \citenamefont {Uchida}, \citenamefont {Noguchi}, \citenamefont {Haze},\ and\
  \citenamefont {Urabe}}]{Toyoda2013}%
  \BibitemOpen
  \bibfield  {author} {\bibinfo {author} {\bibfnamefont {K.}~\bibnamefont
  {Toyoda}}, \bibinfo {author} {\bibfnamefont {K.}~\bibnamefont {Uchida}},
  \bibinfo {author} {\bibfnamefont {A.}~\bibnamefont {Noguchi}}, \bibinfo
  {author} {\bibfnamefont {S.}~\bibnamefont {Haze}},\ and\ \bibinfo {author}
  {\bibfnamefont {S.}~\bibnamefont {Urabe}},\ }\bibfield  {title} {\bibinfo
  {title} {{Realization of holonomic single-qubit operations}},\ }\href
  {https://doi.org/10.1103/PhysRevA.87.052307} {\bibfield  {journal} {\bibinfo
  {journal} {Physical Review A - Atomic, Molecular, and Optical Physics}\
  }\textbf {\bibinfo {volume} {87}},\ \bibinfo {pages} {1} (\bibinfo {year}
  {2013})}\BibitemShut {NoStop}%
\bibitem [{\citenamefont {Wang}\ and\ \citenamefont {Clerk}(2012)}]{Wang2012a}%
  \BibitemOpen
  \bibfield  {author} {\bibinfo {author} {\bibfnamefont {Y.~D.}\ \bibnamefont
  {Wang}}\ and\ \bibinfo {author} {\bibfnamefont {A.~A.}\ \bibnamefont
  {Clerk}},\ }\bibfield  {title} {\bibinfo {title} {{Using interference for
  high fidelity quantum state transfer in optomechanics}},\ }\href
  {https://doi.org/10.1103/PhysRevLett.108.153603} {\bibfield  {journal}
  {\bibinfo  {journal} {Physical Review Letters}\ }\textbf {\bibinfo {volume}
  {108}},\ \bibinfo {pages} {1} (\bibinfo {year} {2012})}\BibitemShut {NoStop}%
\bibitem [{\citenamefont {Tian}(2012)}]{Tian2012}%
  \BibitemOpen
  \bibfield  {author} {\bibinfo {author} {\bibfnamefont {L.}~\bibnamefont
  {Tian}},\ }\bibfield  {title} {\bibinfo {title} {{Adiabatic state conversion
  and pulse transmission in optomechanical systems}},\ }\href
  {https://doi.org/10.1103/PhysRevLett.108.153604} {\bibfield  {journal}
  {\bibinfo  {journal} {Physical Review Letters}\ }\textbf {\bibinfo {volume}
  {108}},\ \bibinfo {pages} {1} (\bibinfo {year} {2012})}\BibitemShut {NoStop}%
\bibitem [{\citenamefont {Garg}\ \emph {et~al.}(2017)\citenamefont {Garg},
  \citenamefont {Chauhan},\ and\ \citenamefont {Biswas}}]{Garg2017}%
  \BibitemOpen
  \bibfield  {author} {\bibinfo {author} {\bibfnamefont {D.}~\bibnamefont
  {Garg}}, \bibinfo {author} {\bibfnamefont {A.~K.}\ \bibnamefont {Chauhan}},\
  and\ \bibinfo {author} {\bibfnamefont {A.}~\bibnamefont {Biswas}},\
  }\bibfield  {title} {\bibinfo {title} {{Adiabatic transfer of energy
  fluctuations between membranes inside an optical cavity}},\ }\bibfield
  {journal} {\bibinfo  {journal} {Physical Review A}\ }\textbf {\bibinfo
  {volume} {96}},\ \href {https://doi.org/10.1103/PhysRevA.96.023837}
  {10.1103/PhysRevA.96.023837} (\bibinfo {year} {2017}),\ \Eprint
  {https://arxiv.org/abs/1706.05869} {arXiv:1706.05869} \BibitemShut {NoStop}%
\bibitem [{\citenamefont {Weaver}\ \emph {et~al.}(2017)\citenamefont {Weaver},
  \citenamefont {Buters}, \citenamefont {Luna}, \citenamefont {Eerkens},
  \citenamefont {Heeck}, \citenamefont {{De Man}},\ and\ \citenamefont
  {Bouwmeester}}]{Weaver2017}%
  \BibitemOpen
  \bibfield  {author} {\bibinfo {author} {\bibfnamefont {M.~J.}\ \bibnamefont
  {Weaver}}, \bibinfo {author} {\bibfnamefont {F.}~\bibnamefont {Buters}},
  \bibinfo {author} {\bibfnamefont {F.}~\bibnamefont {Luna}}, \bibinfo {author}
  {\bibfnamefont {H.}~\bibnamefont {Eerkens}}, \bibinfo {author} {\bibfnamefont
  {K.}~\bibnamefont {Heeck}}, \bibinfo {author} {\bibfnamefont
  {S.}~\bibnamefont {{De Man}}},\ and\ \bibinfo {author} {\bibfnamefont
  {D.}~\bibnamefont {Bouwmeester}},\ }\bibfield  {title} {\bibinfo {title}
  {{Coherent optomechanical state transfer between disparate mechanical
  resonators}},\ }\href {https://doi.org/10.1038/s41467-017-00968-9} {\bibfield
   {journal} {\bibinfo  {journal} {Nature Communications}\ }\textbf {\bibinfo
  {volume} {8}},\ \bibinfo {pages} {1} (\bibinfo {year} {2017})}\BibitemShut
  {NoStop}%
\bibitem [{\citenamefont {Xu}\ \emph {et~al.}(2019)\citenamefont {Xu},
  \citenamefont {Jiang}, \citenamefont {Clerk},\ and\ \citenamefont
  {Harris}}]{Xu2019}%
  \BibitemOpen
  \bibfield  {author} {\bibinfo {author} {\bibfnamefont {H.}~\bibnamefont
  {Xu}}, \bibinfo {author} {\bibfnamefont {L.}~\bibnamefont {Jiang}}, \bibinfo
  {author} {\bibfnamefont {A.~A.}\ \bibnamefont {Clerk}},\ and\ \bibinfo
  {author} {\bibfnamefont {J.~G.}\ \bibnamefont {Harris}},\ }\bibfield  {title}
  {\bibinfo {title} {{Nonreciprocal control and cooling of phonon modes in an
  optomechanical system}},\ }\href {https://doi.org/10.1038/s41586-019-1061-2}
  {\bibfield  {journal} {\bibinfo  {journal} {Nature}\ }\textbf {\bibinfo
  {volume} {568}},\ \bibinfo {pages} {65} (\bibinfo {year} {2019})}\BibitemShut
  {NoStop}%
\bibitem [{\citenamefont {Aspelmeyer}\ \emph {et~al.}(2014)\citenamefont
  {Aspelmeyer}, \citenamefont {Kippenberg},\ and\ \citenamefont
  {Marquardt}}]{Aspelmeyer2014}%
  \BibitemOpen
  \bibfield  {author} {\bibinfo {author} {\bibfnamefont {M.}~\bibnamefont
  {Aspelmeyer}}, \bibinfo {author} {\bibfnamefont {T.~J.}\ \bibnamefont
  {Kippenberg}},\ and\ \bibinfo {author} {\bibfnamefont {F.}~\bibnamefont
  {Marquardt}},\ }\bibfield  {title} {\bibinfo {title} {{Cavity
  optomechanics}},\ }\href {https://doi.org/10.1103/RevModPhys.86.1391}
  {\bibfield  {journal} {\bibinfo  {journal} {Reviews of Modern Physics}\
  }\textbf {\bibinfo {volume} {86}},\ \bibinfo {pages} {1391} (\bibinfo {year}
  {2014})}\BibitemShut {NoStop}%
\bibitem [{\citenamefont {Buchmann}\ and\ \citenamefont
  {Stamper-Kurn}(2015)}]{Buchmann2015}%
  \BibitemOpen
  \bibfield  {author} {\bibinfo {author} {\bibfnamefont {L.~F.}\ \bibnamefont
  {Buchmann}}\ and\ \bibinfo {author} {\bibfnamefont {D.~M.}\ \bibnamefont
  {Stamper-Kurn}},\ }\bibfield  {title} {\bibinfo {title} {{Nondegenerate
  multimode optomechanics}},\ }\bibfield  {journal} {\bibinfo  {journal}
  {Physical Review A - Atomic, Molecular, and Optical Physics}\ }\textbf
  {\bibinfo {volume} {92}},\ \href {https://doi.org/10.1103/PhysRevA.92.013851}
  {10.1103/PhysRevA.92.013851} (\bibinfo {year} {2015})\BibitemShut {NoStop}%
\bibitem [{\citenamefont {Dong}\ \emph {et~al.}(2012)\citenamefont {Dong},
  \citenamefont {Fiore}, \citenamefont {Kuzyk},\ and\ \citenamefont
  {Wang}}]{Dong2012}%
  \BibitemOpen
  \bibfield  {author} {\bibinfo {author} {\bibfnamefont {C.}~\bibnamefont
  {Dong}}, \bibinfo {author} {\bibfnamefont {V.}~\bibnamefont {Fiore}},
  \bibinfo {author} {\bibfnamefont {M.~C.}\ \bibnamefont {Kuzyk}},\ and\
  \bibinfo {author} {\bibfnamefont {H.}~\bibnamefont {Wang}},\ }\bibfield
  {title} {\bibinfo {title} {{Optomechanical Dark Mode}},\ }\href
  {https://doi.org/10.1126/science.1228370} {\bibfield  {journal} {\bibinfo
  {journal} {Science}\ }\textbf {\bibinfo {volume} {338}},\ \bibinfo {pages}
  {1609} (\bibinfo {year} {2012})}\BibitemShut {NoStop}%
\bibitem [{\citenamefont {Thompson}\ \emph {et~al.}(2008)\citenamefont
  {Thompson}, \citenamefont {Zwickl}, \citenamefont {Jayich}, \citenamefont
  {Marquardt}, \citenamefont {Girvin},\ and\ \citenamefont
  {Harris}}]{Thompson2008}%
  \BibitemOpen
  \bibfield  {author} {\bibinfo {author} {\bibfnamefont {J.~D.}\ \bibnamefont
  {Thompson}}, \bibinfo {author} {\bibfnamefont {B.~M.}\ \bibnamefont
  {Zwickl}}, \bibinfo {author} {\bibfnamefont {A.~M.}\ \bibnamefont {Jayich}},
  \bibinfo {author} {\bibfnamefont {F.}~\bibnamefont {Marquardt}}, \bibinfo
  {author} {\bibfnamefont {S.~M.}\ \bibnamefont {Girvin}},\ and\ \bibinfo
  {author} {\bibfnamefont {J.~G.}\ \bibnamefont {Harris}},\ }\bibfield  {title}
  {\bibinfo {title} {{Strong dispersive coupling of a high-finesse cavity to a
  micromechanical membrane}},\ }\href {https://doi.org/10.1038/nature06715}
  {\bibfield  {journal} {\bibinfo  {journal} {Nature}\ }\textbf {\bibinfo
  {volume} {452}},\ \bibinfo {pages} {72} (\bibinfo {year} {2008})}\BibitemShut
  {NoStop}%
\bibitem [{\citenamefont {Tsaturyan}\ \emph {et~al.}(2017)\citenamefont
  {Tsaturyan}, \citenamefont {Barg}, \citenamefont {Polzik},\ and\
  \citenamefont {Schliesser}}]{Tsaturyan2017}%
  \BibitemOpen
  \bibfield  {author} {\bibinfo {author} {\bibfnamefont {Y.}~\bibnamefont
  {Tsaturyan}}, \bibinfo {author} {\bibfnamefont {A.}~\bibnamefont {Barg}},
  \bibinfo {author} {\bibfnamefont {E.~S.}\ \bibnamefont {Polzik}},\ and\
  \bibinfo {author} {\bibfnamefont {A.}~\bibnamefont {Schliesser}},\ }\bibfield
   {title} {\bibinfo {title} {{Ultracoherent nanomechanical resonators via soft
  clamping and dissipation dilution}},\ }\href
  {https://doi.org/10.1038/nnano.2017.101} {\bibfield  {journal} {\bibinfo
  {journal} {Nature Nanotechnology}\ }\textbf {\bibinfo {volume} {12}},\
  \bibinfo {pages} {776} (\bibinfo {year} {2017})}\BibitemShut {NoStop}%
\bibitem [{\citenamefont {Schmid}\ \emph {et~al.}(2011)\citenamefont {Schmid},
  \citenamefont {Jensen}, \citenamefont {Nielsen},\ and\ \citenamefont
  {Boisen}}]{Schmid2011}%
  \BibitemOpen
  \bibfield  {author} {\bibinfo {author} {\bibfnamefont {S.}~\bibnamefont
  {Schmid}}, \bibinfo {author} {\bibfnamefont {K.~D.}\ \bibnamefont {Jensen}},
  \bibinfo {author} {\bibfnamefont {K.~H.}\ \bibnamefont {Nielsen}},\ and\
  \bibinfo {author} {\bibfnamefont {A.}~\bibnamefont {Boisen}},\ }\bibfield
  {title} {\bibinfo {title} {{Damping mechanisms in high-Q micro and
  nanomechanical string resonators}},\ }\href
  {https://doi.org/10.1103/PhysRevB.84.165307} {\bibfield  {journal} {\bibinfo
  {journal} {Physical Review B - Condensed Matter and Materials Physics}\
  }\textbf {\bibinfo {volume} {84}},\ \bibinfo {pages} {1} (\bibinfo {year}
  {2011})}\BibitemShut {NoStop}%
\bibitem [{\citenamefont {Vitanov}\ \emph {et~al.}(2017)\citenamefont
  {Vitanov}, \citenamefont {Rangelov}, \citenamefont {Shore},\ and\
  \citenamefont {Bergmann}}]{Vitanov2017}%
  \BibitemOpen
  \bibfield  {author} {\bibinfo {author} {\bibfnamefont {N.~V.}\ \bibnamefont
  {Vitanov}}, \bibinfo {author} {\bibfnamefont {A.~A.}\ \bibnamefont
  {Rangelov}}, \bibinfo {author} {\bibfnamefont {B.~W.}\ \bibnamefont
  {Shore}},\ and\ \bibinfo {author} {\bibfnamefont {K.}~\bibnamefont
  {Bergmann}},\ }\bibfield  {title} {\bibinfo {title} {{Stimulated Raman
  adiabatic passage in physics, chemistry, and beyond}},\ }\href
  {https://doi.org/10.1103/RevModPhys.89.015006} {\bibfield  {journal}
  {\bibinfo  {journal} {Reviews of Modern Physics}\ }\textbf {\bibinfo {volume}
  {89}},\ \bibinfo {pages} {1} (\bibinfo {year} {2017})}\BibitemShut {NoStop}%
\bibitem [{\citenamefont {Kuhn}\ \emph {et~al.}(1992)\citenamefont {Kuhn},
  \citenamefont {Coulston}, \citenamefont {He}, \citenamefont {Schiemann},
  \citenamefont {Bergmann},\ and\ \citenamefont {Warren}}]{Kuhn1992}%
  \BibitemOpen
  \bibfield  {author} {\bibinfo {author} {\bibfnamefont {A.}~\bibnamefont
  {Kuhn}}, \bibinfo {author} {\bibfnamefont {G.~W.}\ \bibnamefont {Coulston}},
  \bibinfo {author} {\bibfnamefont {G.~Z.}\ \bibnamefont {He}}, \bibinfo
  {author} {\bibfnamefont {S.}~\bibnamefont {Schiemann}}, \bibinfo {author}
  {\bibfnamefont {K.}~\bibnamefont {Bergmann}},\ and\ \bibinfo {author}
  {\bibfnamefont {W.~S.}\ \bibnamefont {Warren}},\ }\bibfield  {title}
  {\bibinfo {title} {{Population transfer by stimulated Raman scattering with
  delayed pulses using spectrally broad light}},\ }\href
  {https://doi.org/10.1063/1.462840} {\bibfield  {journal} {\bibinfo  {journal}
  {The Journal of Chemical Physics}\ }\textbf {\bibinfo {volume} {96}},\
  \bibinfo {pages} {4215} (\bibinfo {year} {1992})}\BibitemShut {NoStop}%
\bibitem [{\citenamefont {Dupont-Nivet}\ \emph {et~al.}(2015)\citenamefont
  {Dupont-Nivet}, \citenamefont {Casiulis}, \citenamefont {Laudat},
  \citenamefont {Westbrook},\ and\ \citenamefont
  {Schwartz}}]{Dupont-Nivet2015}%
  \BibitemOpen
  \bibfield  {author} {\bibinfo {author} {\bibfnamefont {M.}~\bibnamefont
  {Dupont-Nivet}}, \bibinfo {author} {\bibfnamefont {M.}~\bibnamefont
  {Casiulis}}, \bibinfo {author} {\bibfnamefont {T.}~\bibnamefont {Laudat}},
  \bibinfo {author} {\bibfnamefont {C.~I.}\ \bibnamefont {Westbrook}},\ and\
  \bibinfo {author} {\bibfnamefont {S.}~\bibnamefont {Schwartz}},\ }\bibfield
  {title} {\bibinfo {title} {{Microwave-stimulated Raman adiabatic passage in a
  Bose-Einstein condensate on an atom chip}},\ }\href
  {https://doi.org/10.1103/PhysRevA.91.053420} {\bibfield  {journal} {\bibinfo
  {journal} {Physical Review A - Atomic, Molecular, and Optical Physics}\
  }\textbf {\bibinfo {volume} {91}},\ \bibinfo {pages} {1} (\bibinfo {year}
  {2015})},\ \Eprint {https://arxiv.org/abs/1503.05360} {arXiv:1503.05360}
  \BibitemShut {NoStop}%
\bibitem [{\citenamefont {Xu}\ \emph {et~al.}(2016)\citenamefont {Xu},
  \citenamefont {Song}, \citenamefont {Liu}, \citenamefont {Xue}, \citenamefont
  {Su}, \citenamefont {Deng}, \citenamefont {Tian}, \citenamefont {Zheng},
  \citenamefont {Han}, \citenamefont {Zhong}, \citenamefont {Wang},
  \citenamefont {Liu},\ and\ \citenamefont {Zhao}}]{Xu2016a}%
  \BibitemOpen
  \bibfield  {author} {\bibinfo {author} {\bibfnamefont {H.~K.}\ \bibnamefont
  {Xu}}, \bibinfo {author} {\bibfnamefont {C.}~\bibnamefont {Song}}, \bibinfo
  {author} {\bibfnamefont {W.~Y.}\ \bibnamefont {Liu}}, \bibinfo {author}
  {\bibfnamefont {G.~M.}\ \bibnamefont {Xue}}, \bibinfo {author} {\bibfnamefont
  {F.~F.}\ \bibnamefont {Su}}, \bibinfo {author} {\bibfnamefont
  {H.}~\bibnamefont {Deng}}, \bibinfo {author} {\bibfnamefont {Y.}~\bibnamefont
  {Tian}}, \bibinfo {author} {\bibfnamefont {D.~N.}\ \bibnamefont {Zheng}},
  \bibinfo {author} {\bibfnamefont {S.}~\bibnamefont {Han}}, \bibinfo {author}
  {\bibfnamefont {Y.~P.}\ \bibnamefont {Zhong}}, \bibinfo {author}
  {\bibfnamefont {H.}~\bibnamefont {Wang}}, \bibinfo {author} {\bibfnamefont
  {Y.~X.}\ \bibnamefont {Liu}},\ and\ \bibinfo {author} {\bibfnamefont {S.~P.}\
  \bibnamefont {Zhao}},\ }\bibfield  {title} {\bibinfo {title} {{Coherent
  population transfer between uncoupled or weakly coupled states in ladder-type
  superconducting qutrits}},\ }\href {https://doi.org/10.1038/ncomms11018}
  {\bibfield  {journal} {\bibinfo  {journal} {Nature Communications}\ }\textbf
  {\bibinfo {volume} {7}},\ \bibinfo {pages} {1} (\bibinfo {year}
  {2016})}\BibitemShut {NoStop}%
\bibitem [{\citenamefont {Klein}\ \emph {et~al.}(2008)\citenamefont {Klein},
  \citenamefont {Beil},\ and\ \citenamefont {Halfmann}}]{Klein2008}%
  \BibitemOpen
  \bibfield  {author} {\bibinfo {author} {\bibfnamefont {J.}~\bibnamefont
  {Klein}}, \bibinfo {author} {\bibfnamefont {F.}~\bibnamefont {Beil}},\ and\
  \bibinfo {author} {\bibfnamefont {T.}~\bibnamefont {Halfmann}},\ }\bibfield
  {title} {\bibinfo {title} {{Experimental investigations of stimulated Raman
  adiabatic passage in a doped solid}},\ }\href
  {https://doi.org/10.1103/PhysRevA.78.033416} {\bibfield  {journal} {\bibinfo
  {journal} {Physical Review A - Atomic, Molecular, and Optical Physics}\
  }\textbf {\bibinfo {volume} {78}},\ \bibinfo {pages} {1} (\bibinfo {year}
  {2008})}\BibitemShut {NoStop}%
\bibitem [{\citenamefont {Theuer}\ and\ \citenamefont
  {Bergmann}(1998)}]{Theuer1998}%
  \BibitemOpen
  \bibfield  {author} {\bibinfo {author} {\bibfnamefont {H.}~\bibnamefont
  {Theuer}}\ and\ \bibinfo {author} {\bibfnamefont {K.}~\bibnamefont
  {Bergmann}},\ }\bibfield  {title} {\bibinfo {title} {{Atomic beam deflection
  by coherent momentum transfer and the dependence on weak magnetic fields}},\
  }\href {https://doi.org/10.1007/s100530050141} {\bibfield  {journal}
  {\bibinfo  {journal} {European Physical Journal D}\ }\textbf {\bibinfo
  {volume} {2}},\ \bibinfo {pages} {279} (\bibinfo {year} {1998})}\BibitemShut
  {NoStop}%
\bibitem [{\citenamefont {Rossi}\ \emph {et~al.}(2018)\citenamefont {Rossi},
  \citenamefont {Mason}, \citenamefont {Chen}, \citenamefont {Tsaturyan},\ and\
  \citenamefont {Schliesser}}]{Rossi2018}%
  \BibitemOpen
  \bibfield  {author} {\bibinfo {author} {\bibfnamefont {M.}~\bibnamefont
  {Rossi}}, \bibinfo {author} {\bibfnamefont {D.}~\bibnamefont {Mason}},
  \bibinfo {author} {\bibfnamefont {J.}~\bibnamefont {Chen}}, \bibinfo {author}
  {\bibfnamefont {Y.}~\bibnamefont {Tsaturyan}},\ and\ \bibinfo {author}
  {\bibfnamefont {A.}~\bibnamefont {Schliesser}},\ }\bibfield  {title}
  {\bibinfo {title} {{Measurement-based quantum control of mechanical
  motion}},\ }\href {https://doi.org/10.1038/s41586-018-0643-8} {\bibfield
  {journal} {\bibinfo  {journal} {Nature}\ }\textbf {\bibinfo {volume} {563}},\
  \bibinfo {pages} {53} (\bibinfo {year} {2018})},\ \Eprint
  {https://arxiv.org/abs/1805.05087} {arXiv:1805.05087} \BibitemShut {NoStop}%
\bibitem [{\citenamefont {Peterson}\ \emph {et~al.}(2016)\citenamefont
  {Peterson}, \citenamefont {Purdy}, \citenamefont {Kampel}, \citenamefont
  {Andrews}, \citenamefont {Yu}, \citenamefont {Lehnert},\ and\ \citenamefont
  {Regal}}]{Peterson2016}%
  \BibitemOpen
  \bibfield  {author} {\bibinfo {author} {\bibfnamefont {R.~W.}\ \bibnamefont
  {Peterson}}, \bibinfo {author} {\bibfnamefont {T.~P.}\ \bibnamefont {Purdy}},
  \bibinfo {author} {\bibfnamefont {N.~S.}\ \bibnamefont {Kampel}}, \bibinfo
  {author} {\bibfnamefont {R.~W.}\ \bibnamefont {Andrews}}, \bibinfo {author}
  {\bibfnamefont {P.~L.}\ \bibnamefont {Yu}}, \bibinfo {author} {\bibfnamefont
  {K.~W.}\ \bibnamefont {Lehnert}},\ and\ \bibinfo {author} {\bibfnamefont
  {C.~A.}\ \bibnamefont {Regal}},\ }\bibfield  {title} {\bibinfo {title}
  {{Laser Cooling of a Micromechanical Membrane to the Quantum Backaction
  Limit}},\ }\href {https://doi.org/10.1103/PhysRevLett.116.063601} {\bibfield
  {journal} {\bibinfo  {journal} {Physical Review Letters}\ }\textbf {\bibinfo
  {volume} {116}},\ \bibinfo {pages} {1} (\bibinfo {year} {2016})}\BibitemShut
  {NoStop}%
\bibitem [{\citenamefont {Underwood}\ \emph {et~al.}(2015)\citenamefont
  {Underwood}, \citenamefont {Mason}, \citenamefont {Lee}, \citenamefont {Xu},
  \citenamefont {Jiang}, \citenamefont {Shkarin}, \citenamefont {B{\o}rkje},
  \citenamefont {Girvin},\ and\ \citenamefont {Harris}}]{Underwood2015}%
  \BibitemOpen
  \bibfield  {author} {\bibinfo {author} {\bibfnamefont {M.}~\bibnamefont
  {Underwood}}, \bibinfo {author} {\bibfnamefont {D.}~\bibnamefont {Mason}},
  \bibinfo {author} {\bibfnamefont {D.}~\bibnamefont {Lee}}, \bibinfo {author}
  {\bibfnamefont {H.}~\bibnamefont {Xu}}, \bibinfo {author} {\bibfnamefont
  {L.}~\bibnamefont {Jiang}}, \bibinfo {author} {\bibfnamefont {A.~B.}\
  \bibnamefont {Shkarin}}, \bibinfo {author} {\bibfnamefont {K.}~\bibnamefont
  {B{\o}rkje}}, \bibinfo {author} {\bibfnamefont {S.~M.}\ \bibnamefont
  {Girvin}},\ and\ \bibinfo {author} {\bibfnamefont {J.~G.}\ \bibnamefont
  {Harris}},\ }\bibfield  {title} {\bibinfo {title} {{Measurement of the
  motional sidebands of a nanogram-scale oscillator in the quantum regime}},\
  }\href {https://doi.org/10.1103/PhysRevA.92.061801} {\bibfield  {journal}
  {\bibinfo  {journal} {Physical Review A - Atomic, Molecular, and Optical
  Physics}\ }\textbf {\bibinfo {volume} {92}},\ \bibinfo {pages} {1} (\bibinfo
  {year} {2015})}\BibitemShut {NoStop}%
\bibitem [{\citenamefont {Marte}\ \emph {et~al.}(1991)\citenamefont {Marte},
  \citenamefont {Zoller},\ and\ \citenamefont {Hall}}]{Marte1991}%
  \BibitemOpen
  \bibfield  {author} {\bibinfo {author} {\bibfnamefont {P.}~\bibnamefont
  {Marte}}, \bibinfo {author} {\bibfnamefont {P.}~\bibnamefont {Zoller}},\ and\
  \bibinfo {author} {\bibfnamefont {J.~L.}\ \bibnamefont {Hall}},\ }\bibfield
  {title} {\bibinfo {title} {{Coherent atomic mirrors and beam splitters by
  adiabatic passage in multilevel systems}},\ }\href
  {https://doi.org/10.1103/PhysRevA.44.R4118} {\bibfield  {journal} {\bibinfo
  {journal} {Physical Review A}\ }\textbf {\bibinfo {volume} {44}},\ \bibinfo
  {pages} {4118} (\bibinfo {year} {1991})}\BibitemShut {NoStop}%
\bibitem [{\citenamefont {Unanyan}\ \emph {et~al.}(1998)\citenamefont
  {Unanyan}, \citenamefont {Fleischhauer}, \citenamefont {Shore},\ and\
  \citenamefont {Bergmann}}]{Unanyan1998}%
  \BibitemOpen
  \bibfield  {author} {\bibinfo {author} {\bibfnamefont {R.}~\bibnamefont
  {Unanyan}}, \bibinfo {author} {\bibfnamefont {M.}~\bibnamefont
  {Fleischhauer}}, \bibinfo {author} {\bibfnamefont {B.~W.}\ \bibnamefont
  {Shore}},\ and\ \bibinfo {author} {\bibfnamefont {K.}~\bibnamefont
  {Bergmann}},\ }\bibfield  {title} {\bibinfo {title} {{Robust creation and
  phase-sensitive probing of superposition states via stimulated Raman
  adiabatic passage (STIRAP) with degenerate dark states}},\ }\href
  {https://doi.org/10.1016/S0030-4018(98)00358-7} {\bibfield  {journal}
  {\bibinfo  {journal} {Optics Communications}\ }\textbf {\bibinfo {volume}
  {155}},\ \bibinfo {pages} {144} (\bibinfo {year} {1998})}\BibitemShut
  {NoStop}%
\end{thebibliography}%


\begin{thebibliography}{17}%
\makeatletter
\providecommand \@ifxundefined [1]{%
 \@ifx{#1\undefined}
}%
\providecommand \@ifnum [1]{%
 \ifnum #1\expandafter \@firstoftwo
 \else \expandafter \@secondoftwo
 \fi
}%
\providecommand \@ifx [1]{%
 \ifx #1\expandafter \@firstoftwo
 \else \expandafter \@secondoftwo
 \fi
}%
\providecommand \natexlab [1]{#1}%
\providecommand \enquote  [1]{``#1''}%
\providecommand \bibnamefont  [1]{#1}%
\providecommand \bibfnamefont [1]{#1}%
\providecommand \citenamefont [1]{#1}%
\providecommand \href@noop [0]{\@secondoftwo}%
\providecommand \href [0]{\begingroup \@sanitize@url \@href}%
\providecommand \@href[1]{\@@startlink{#1}\@@href}%
\providecommand \@@href[1]{\endgroup#1\@@endlink}%
\providecommand \@sanitize@url [0]{\catcode `\\12\catcode `\$12\catcode
  `\&12\catcode `\#12\catcode `\^12\catcode `\_12\catcode `\%12\relax}%
\providecommand \@@startlink[1]{}%
\providecommand \@@endlink[0]{}%
\providecommand \url  [0]{\begingroup\@sanitize@url \@url }%
\providecommand \@url [1]{\endgroup\@href {#1}{\urlprefix }}%
\providecommand \urlprefix  [0]{URL }%
\providecommand \Eprint [0]{\href }%
\providecommand \doibase [0]{https://doi.org/}%
\providecommand \selectlanguage [0]{\@gobble}%
\providecommand \bibinfo  [0]{\@secondoftwo}%
\providecommand \bibfield  [0]{\@secondoftwo}%
\providecommand \translation [1]{[#1]}%
\providecommand \BibitemOpen [0]{}%
\providecommand \bibitemStop [0]{}%
\providecommand \bibitemNoStop [0]{.\EOS\space}%
\providecommand \EOS [0]{\spacefactor3000\relax}%
\providecommand \BibitemShut  [1]{\csname bibitem#1\endcsname}%
\let\auto@bib@innerbib\@empty
\bibitem [{\citenamefont {Drever}\ \emph {et~al.}(1983)\citenamefont {Drever},
  \citenamefont {Hall}, \citenamefont {Kowalski}, \citenamefont {Hough},
  \citenamefont {Ford}, \citenamefont {Munley},\ and\ \citenamefont
  {Ward}}]{Drever1983}%
  \BibitemOpen
  \bibfield  {author} {\bibinfo {author} {\bibfnamefont {R.~W.}\ \bibnamefont
  {Drever}}, \bibinfo {author} {\bibfnamefont {J.~L.}\ \bibnamefont {Hall}},
  \bibinfo {author} {\bibfnamefont {F.~V.}\ \bibnamefont {Kowalski}}, \bibinfo
  {author} {\bibfnamefont {J.}~\bibnamefont {Hough}}, \bibinfo {author}
  {\bibfnamefont {G.~M.}\ \bibnamefont {Ford}}, \bibinfo {author}
  {\bibfnamefont {A.~J.}\ \bibnamefont {Munley}},\ and\ \bibinfo {author}
  {\bibfnamefont {H.}~\bibnamefont {Ward}},\ }\bibfield  {title} {\bibinfo
  {title} {{Laser phase and frequency stabilization using an optical
  resonator}},\ }\href {https://doi.org/10.1007/BF00702605} {\bibfield
  {journal} {\bibinfo  {journal} {Applied Physics B Photophysics and Laser
  Chemistry}\ }\textbf {\bibinfo {volume} {31}},\ \bibinfo {pages} {97}
  (\bibinfo {year} {1983})}\BibitemShut {NoStop}%
\bibitem [{\citenamefont {Jayich}\ \emph {et~al.}(2008)\citenamefont {Jayich},
  \citenamefont {Sankey}, \citenamefont {Zwickl}, \citenamefont {Yang},
  \citenamefont {Thompson}, \citenamefont {Girvin}, \citenamefont {Clerk},
  \citenamefont {Marquardt},\ and\ \citenamefont {Harris}}]{Jayich2008}%
  \BibitemOpen
  \bibfield  {author} {\bibinfo {author} {\bibfnamefont {A.~M.}\ \bibnamefont
  {Jayich}}, \bibinfo {author} {\bibfnamefont {J.~C.}\ \bibnamefont {Sankey}},
  \bibinfo {author} {\bibfnamefont {B.~M.}\ \bibnamefont {Zwickl}}, \bibinfo
  {author} {\bibfnamefont {C.}~\bibnamefont {Yang}}, \bibinfo {author}
  {\bibfnamefont {J.~D.}\ \bibnamefont {Thompson}}, \bibinfo {author}
  {\bibfnamefont {S.~M.}\ \bibnamefont {Girvin}}, \bibinfo {author}
  {\bibfnamefont {A.~A.}\ \bibnamefont {Clerk}}, \bibinfo {author}
  {\bibfnamefont {F.}~\bibnamefont {Marquardt}},\ and\ \bibinfo {author}
  {\bibfnamefont {J.~G.}\ \bibnamefont {Harris}},\ }\bibfield  {title}
  {\bibinfo {title} {{Dispersive optomechanics: A membrane inside a cavity}},\
  }\bibfield  {journal} {\bibinfo  {journal} {New Journal of Physics}\ }\textbf
  {\bibinfo {volume} {10}},\ \href
  {https://doi.org/10.1088/1367-2630/10/9/095008}
  {10.1088/1367-2630/10/9/095008} (\bibinfo {year} {2008}),\ \Eprint
  {https://arxiv.org/abs/arXiv:0805.3723v1} {arXiv:arXiv:0805.3723v1}
  \BibitemShut {NoStop}%
\bibitem [{\citenamefont {Aspelmeyer}\ \emph {et~al.}(2014)\citenamefont
  {Aspelmeyer}, \citenamefont {Kippenberg},\ and\ \citenamefont
  {Marquardt}}]{Aspelmeyer2014}%
  \BibitemOpen
  \bibfield  {author} {\bibinfo {author} {\bibfnamefont {M.}~\bibnamefont
  {Aspelmeyer}}, \bibinfo {author} {\bibfnamefont {T.~J.}\ \bibnamefont
  {Kippenberg}},\ and\ \bibinfo {author} {\bibfnamefont {F.}~\bibnamefont
  {Marquardt}},\ }\bibfield  {title} {\bibinfo {title} {{Cavity
  optomechanics}},\ }\href {https://doi.org/10.1103/RevModPhys.86.1391}
  {\bibfield  {journal} {\bibinfo  {journal} {Reviews of Modern Physics}\
  }\textbf {\bibinfo {volume} {86}},\ \bibinfo {pages} {1391} (\bibinfo {year}
  {2014})}\BibitemShut {NoStop}%
\bibitem [{\citenamefont {Peterson}\ \emph {et~al.}(2016)\citenamefont
  {Peterson}, \citenamefont {Purdy}, \citenamefont {Kampel}, \citenamefont
  {Andrews}, \citenamefont {Yu}, \citenamefont {Lehnert},\ and\ \citenamefont
  {Regal}}]{Laser_Cooling}%
  \BibitemOpen
  \bibfield  {author} {\bibinfo {author} {\bibfnamefont {R.~W.}\ \bibnamefont
  {Peterson}}, \bibinfo {author} {\bibfnamefont {T.~P.}\ \bibnamefont {Purdy}},
  \bibinfo {author} {\bibfnamefont {N.~S.}\ \bibnamefont {Kampel}}, \bibinfo
  {author} {\bibfnamefont {R.~W.}\ \bibnamefont {Andrews}}, \bibinfo {author}
  {\bibfnamefont {P.-L.}\ \bibnamefont {Yu}}, \bibinfo {author} {\bibfnamefont
  {K.~W.}\ \bibnamefont {Lehnert}},\ and\ \bibinfo {author} {\bibfnamefont
  {C.~A.}\ \bibnamefont {Regal}},\ }\bibfield  {title} {\bibinfo {title} {Laser
  cooling of a micromechanical membrane to the quantum backaction limit},\
  }\href {https://doi.org/10.1103/PhysRevLett.116.063601} {\bibfield  {journal}
  {\bibinfo  {journal} {Phys. Rev. Lett.}\ }\textbf {\bibinfo {volume} {116}},\
  \bibinfo {pages} {063601} (\bibinfo {year} {2016})}\BibitemShut {NoStop}%
\bibitem [{\citenamefont {Underwood}\ \emph {et~al.}(2015)\citenamefont
  {Underwood}, \citenamefont {Mason}, \citenamefont {Lee}, \citenamefont {Xu},
  \citenamefont {Jiang}, \citenamefont {Shkarin}, \citenamefont {B\o{}rkje},
  \citenamefont {Girvin},\ and\ \citenamefont {Harris}}]{Measurement_Motional}%
  \BibitemOpen
  \bibfield  {author} {\bibinfo {author} {\bibfnamefont {M.}~\bibnamefont
  {Underwood}}, \bibinfo {author} {\bibfnamefont {D.}~\bibnamefont {Mason}},
  \bibinfo {author} {\bibfnamefont {D.}~\bibnamefont {Lee}}, \bibinfo {author}
  {\bibfnamefont {H.}~\bibnamefont {Xu}}, \bibinfo {author} {\bibfnamefont
  {L.}~\bibnamefont {Jiang}}, \bibinfo {author} {\bibfnamefont {A.~B.}\
  \bibnamefont {Shkarin}}, \bibinfo {author} {\bibfnamefont {K.}~\bibnamefont
  {B\o{}rkje}}, \bibinfo {author} {\bibfnamefont {S.~M.}\ \bibnamefont
  {Girvin}},\ and\ \bibinfo {author} {\bibfnamefont {J.~G.~E.}\ \bibnamefont
  {Harris}},\ }\bibfield  {title} {\bibinfo {title} {Measurement of the
  motional sidebands of a nanogram-scale oscillator in the quantum regime},\
  }\href {https://doi.org/10.1103/PhysRevA.92.061801} {\bibfield  {journal}
  {\bibinfo  {journal} {Phys. Rev. A}\ }\textbf {\bibinfo {volume} {92}},\
  \bibinfo {pages} {061801} (\bibinfo {year} {2015})}\BibitemShut {NoStop}%
\bibitem [{\citenamefont {Rossi}\ \emph {et~al.}(2018)\citenamefont {Rossi},
  \citenamefont {Mason}, \citenamefont {Chen}, \citenamefont {Tsaturyan},\ and\
  \citenamefont {Schliesser}}]{Measurement-based}%
  \BibitemOpen
  \bibfield  {author} {\bibinfo {author} {\bibfnamefont {M.}~\bibnamefont
  {Rossi}}, \bibinfo {author} {\bibfnamefont {D.}~\bibnamefont {Mason}},
  \bibinfo {author} {\bibfnamefont {J.}~\bibnamefont {Chen}}, \bibinfo {author}
  {\bibfnamefont {Y.}~\bibnamefont {Tsaturyan}},\ and\ \bibinfo {author}
  {\bibfnamefont {A.}~\bibnamefont {Schliesser}},\ }\bibfield  {title}
  {\bibinfo {title} {Measurement-based quantum control of mechanical motion},\
  }\href {https://doi.org/10.1038/s41586-018-0643-8} {\bibfield  {journal}
  {\bibinfo  {journal} {Nature}\ }\textbf {\bibinfo {volume} {563}},\ \bibinfo
  {pages} {531} (\bibinfo {year} {2018})}\BibitemShut {NoStop}%
\bibitem [{\citenamefont {Galinskiy}\ \emph {et~al.}(2020)\citenamefont
  {Galinskiy}, \citenamefont {Tsaturyan}, \citenamefont {Parniak},\ and\
  \citenamefont {Polzik}}]{Polzik}%
  \BibitemOpen
  \bibfield  {author} {\bibinfo {author} {\bibfnamefont {I.}~\bibnamefont
  {Galinskiy}}, \bibinfo {author} {\bibfnamefont {Y.}~\bibnamefont
  {Tsaturyan}}, \bibinfo {author} {\bibfnamefont {M.}~\bibnamefont {Parniak}},\
  and\ \bibinfo {author} {\bibfnamefont {E.~S.}\ \bibnamefont {Polzik}},\
  }\bibfield  {title} {\bibinfo {title} {Phonon counting thermometry of an
  ultracoherent membrane resonator near its motional ground state},\ }\href
  {https://doi.org/10.1364/OPTICA.390939} {\bibfield  {journal} {\bibinfo
  {journal} {Optica}\ }\textbf {\bibinfo {volume} {7}},\ \bibinfo {pages} {718}
  (\bibinfo {year} {2020})}\BibitemShut {NoStop}%
\bibitem [{\citenamefont {Zhai}\ \emph {et~al.}(2020)\citenamefont {Zhai},
  \citenamefont {Chen},\ and\ \citenamefont {Lin}}]{ground_state_theory}%
  \BibitemOpen
  \bibfield  {author} {\bibinfo {author} {\bibfnamefont {Y.}~\bibnamefont
  {Zhai}}, \bibinfo {author} {\bibfnamefont {Z.~X.}\ \bibnamefont {Chen}},\
  and\ \bibinfo {author} {\bibfnamefont {Q.}~\bibnamefont {Lin}},\ }\bibfield
  {title} {\bibinfo {title} {Efficient ground state cooling of a mechanical
  resonator in a membrane-in-the-middle system by a single drive},\ }\href
  {https://doi.org/10.1364/JOSAB.384108} {\bibfield  {journal} {\bibinfo
  {journal} {J. Opt. Soc. Am. B}\ }\textbf {\bibinfo {volume} {37}},\ \bibinfo
  {pages} {956} (\bibinfo {year} {2020})}\BibitemShut {NoStop}%
\bibitem [{\citenamefont {Flayac}\ and\ \citenamefont
  {Savona}(2014)}]{Heralded_prep}%
  \BibitemOpen
  \bibfield  {author} {\bibinfo {author} {\bibfnamefont {H.}~\bibnamefont
  {Flayac}}\ and\ \bibinfo {author} {\bibfnamefont {V.}~\bibnamefont
  {Savona}},\ }\bibfield  {title} {\bibinfo {title} {Heralded preparation and
  readout of entangled phonons in a photonic crystal cavity},\ }\href
  {https://doi.org/10.1103/PhysRevLett.113.143603} {\bibfield  {journal}
  {\bibinfo  {journal} {Phys. Rev. Lett.}\ }\textbf {\bibinfo {volume} {113}},\
  \bibinfo {pages} {143603} (\bibinfo {year} {2014})}\BibitemShut {NoStop}%
\bibitem [{\citenamefont {Riedinger}\ \emph {et~al.}(2018)\citenamefont
  {Riedinger}, \citenamefont {Wallucks}, \citenamefont {Marinković},
  \citenamefont {Löschnauer}, \citenamefont {Aspelmeyer}, \citenamefont
  {Hong},\ and\ \citenamefont {Gröblacher}}]{Remote}%
  \BibitemOpen
  \bibfield  {author} {\bibinfo {author} {\bibfnamefont {R.}~\bibnamefont
  {Riedinger}}, \bibinfo {author} {\bibfnamefont {A.}~\bibnamefont {Wallucks}},
  \bibinfo {author} {\bibfnamefont {I.}~\bibnamefont {Marinković}}, \bibinfo
  {author} {\bibfnamefont {C.}~\bibnamefont {Löschnauer}}, \bibinfo {author}
  {\bibfnamefont {M.}~\bibnamefont {Aspelmeyer}}, \bibinfo {author}
  {\bibfnamefont {S.}~\bibnamefont {Hong}},\ and\ \bibinfo {author}
  {\bibfnamefont {S.}~\bibnamefont {Gröblacher}},\ }\bibfield  {title}
  {\bibinfo {title} {Remote quantum entanglement between two micromechanical
  oscillators},\ }\href {https://doi.org/10.1038/s41586-018-0036-z} {\bibfield
  {journal} {\bibinfo  {journal} {Nature}\ }\textbf {\bibinfo {volume} {556}},\
  \bibinfo {pages} {473} (\bibinfo {year} {2018})}\BibitemShut {NoStop}%
\bibitem [{\citenamefont {Johansson}\ \emph {et~al.}(2012)\citenamefont
  {Johansson}, \citenamefont {Nation},\ and\ \citenamefont {Nori}}]{QuTiP}%
  \BibitemOpen
  \bibfield  {author} {\bibinfo {author} {\bibfnamefont {J.}~\bibnamefont
  {Johansson}}, \bibinfo {author} {\bibfnamefont {P.}~\bibnamefont {Nation}},\
  and\ \bibinfo {author} {\bibfnamefont {F.}~\bibnamefont {Nori}},\ }\bibfield
  {title} {\bibinfo {title} {Qutip: An open-source python framework for the
  dynamics of open quantum systems},\ }\href
  {https://doi.org/10.1016/j.cpc.2012.02.021} {\bibfield  {journal} {\bibinfo
  {journal} {Computer Physics Communications}\ }\textbf {\bibinfo {volume}
  {183}},\ \bibinfo {pages} {1760–1772} (\bibinfo {year} {2012})}\BibitemShut
  {NoStop}%
\bibitem [{\citenamefont {Buchmann}\ and\ \citenamefont
  {Stamper-Kurn}(2015)}]{multimode}%
  \BibitemOpen
  \bibfield  {author} {\bibinfo {author} {\bibfnamefont {L.~F.}\ \bibnamefont
  {Buchmann}}\ and\ \bibinfo {author} {\bibfnamefont {D.~M.}\ \bibnamefont
  {Stamper-Kurn}},\ }\bibfield  {title} {\bibinfo {title} {Nondegenerate
  multimode optomechanics},\ }\href
  {https://doi.org/10.1103/PhysRevA.92.013851} {\bibfield  {journal} {\bibinfo
  {journal} {Phys. Rev. A}\ }\textbf {\bibinfo {volume} {92}},\ \bibinfo
  {pages} {013851} (\bibinfo {year} {2015})}\BibitemShut {NoStop}%
\bibitem [{\citenamefont {Fleming}\ \emph {et~al.}(2010)\citenamefont
  {Fleming}, \citenamefont {Cummings}, \citenamefont {Anastopoulos},\ and\
  \citenamefont {Hu}}]{Secular}%
  \BibitemOpen
  \bibfield  {author} {\bibinfo {author} {\bibfnamefont {C.}~\bibnamefont
  {Fleming}}, \bibinfo {author} {\bibfnamefont {N.~I.}\ \bibnamefont
  {Cummings}}, \bibinfo {author} {\bibfnamefont {C.}~\bibnamefont
  {Anastopoulos}},\ and\ \bibinfo {author} {\bibfnamefont {B.~L.}\ \bibnamefont
  {Hu}},\ }\bibfield  {title} {\bibinfo {title} {The rotating-wave
  approximation: consistency and applicability from an open quantum system
  analysis},\ }\href {https://doi.org/10.1088/1751-8113/43/40/405304}
  {\bibfield  {journal} {\bibinfo  {journal} {Journal of Physics A:
  Mathematical and Theoretical}\ }\textbf {\bibinfo {volume} {43}},\ \bibinfo
  {pages} {405304} (\bibinfo {year} {2010})}\BibitemShut {NoStop}%
\bibitem [{\citenamefont {Weaver}\ \emph {et~al.}(2018)\citenamefont {Weaver},
  \citenamefont {Newsom}, \citenamefont {Luna}, \citenamefont {L\"offler},\
  and\ \citenamefont {Bouwmeester}}]{interferometry}%
  \BibitemOpen
  \bibfield  {author} {\bibinfo {author} {\bibfnamefont {M.~J.}\ \bibnamefont
  {Weaver}}, \bibinfo {author} {\bibfnamefont {D.}~\bibnamefont {Newsom}},
  \bibinfo {author} {\bibfnamefont {F.}~\bibnamefont {Luna}}, \bibinfo {author}
  {\bibfnamefont {W.}~\bibnamefont {L\"offler}},\ and\ \bibinfo {author}
  {\bibfnamefont {D.}~\bibnamefont {Bouwmeester}},\ }\bibfield  {title}
  {\bibinfo {title} {Phonon interferometry for measuring quantum decoherence},\
  }\href {https://doi.org/10.1103/PhysRevA.97.063832} {\bibfield  {journal}
  {\bibinfo  {journal} {Phys. Rev. A}\ }\textbf {\bibinfo {volume} {97}},\
  \bibinfo {pages} {063832} (\bibinfo {year} {2018})}\BibitemShut {NoStop}%
\bibitem [{\citenamefont {Shibata}\ \emph {et~al.}(2013)\citenamefont
  {Shibata}, \citenamefont {Shimizu}, \citenamefont {Takesue},\ and\
  \citenamefont {Tokura}}]{Superconducting_DCR}%
  \BibitemOpen
  \bibfield  {author} {\bibinfo {author} {\bibfnamefont {H.}~\bibnamefont
  {Shibata}}, \bibinfo {author} {\bibfnamefont {K.}~\bibnamefont {Shimizu}},
  \bibinfo {author} {\bibfnamefont {H.}~\bibnamefont {Takesue}},\ and\ \bibinfo
  {author} {\bibfnamefont {Y.}~\bibnamefont {Tokura}},\ }\bibfield  {title}
  {\bibinfo {title} {Superconducting nanowire single-photon detector with
  ultralow dark count rate using cold optical filters},\ }\href
  {https://doi.org/10.7567/apex.6.072801} {\bibfield  {journal} {\bibinfo
  {journal} {Applied Physics Express}\ }\textbf {\bibinfo {volume} {6}},\
  \bibinfo {pages} {072801} (\bibinfo {year} {2013})}\BibitemShut {NoStop}%
\bibitem [{\citenamefont {Shibata}\ \emph {et~al.}(2015)\citenamefont
  {Shibata}, \citenamefont {Shimizu}, \citenamefont {Takesue},\ and\
  \citenamefont {Tokura}}]{Ultimate_DCR}%
  \BibitemOpen
  \bibfield  {author} {\bibinfo {author} {\bibfnamefont {H.}~\bibnamefont
  {Shibata}}, \bibinfo {author} {\bibfnamefont {K.}~\bibnamefont {Shimizu}},
  \bibinfo {author} {\bibfnamefont {H.}~\bibnamefont {Takesue}},\ and\ \bibinfo
  {author} {\bibfnamefont {Y.}~\bibnamefont {Tokura}},\ }\bibfield  {title}
  {\bibinfo {title} {Ultimate low system dark-count rate for superconducting
  nanowire single-photon detector},\ }\href
  {https://doi.org/10.1364/OL.40.003428} {\bibfield  {journal} {\bibinfo
  {journal} {Opt. Lett.}\ }\textbf {\bibinfo {volume} {40}},\ \bibinfo {pages}
  {3428} (\bibinfo {year} {2015})}\BibitemShut {NoStop}%
\bibitem [{\citenamefont {Vitanov}\ \emph {et~al.}(2017)\citenamefont
  {Vitanov}, \citenamefont {Rangelov}, \citenamefont {Shore},\ and\
  \citenamefont {Bergmann}}]{Vitanov2017}%
  \BibitemOpen
  \bibfield  {author} {\bibinfo {author} {\bibfnamefont {N.~V.}\ \bibnamefont
  {Vitanov}}, \bibinfo {author} {\bibfnamefont {A.~A.}\ \bibnamefont
  {Rangelov}}, \bibinfo {author} {\bibfnamefont {B.~W.}\ \bibnamefont
  {Shore}},\ and\ \bibinfo {author} {\bibfnamefont {K.}~\bibnamefont
  {Bergmann}},\ }\bibfield  {title} {\bibinfo {title} {{Stimulated Raman
  adiabatic passage in physics, chemistry, and beyond}},\ }\href
  {https://doi.org/10.1103/RevModPhys.89.015006} {\bibfield  {journal}
  {\bibinfo  {journal} {Reviews of Modern Physics}\ }\textbf {\bibinfo {volume}
  {89}},\ \bibinfo {pages} {1} (\bibinfo {year} {2017})}\BibitemShut {NoStop}%
\end{thebibliography}%




\end{document}


\title{Supplemental Material: Stimulated Raman adiabatic passage in optomechanics}

\author{Vitaly Fedoseev$^{1}$, Fernando Luna$^2$, Ian Hedgepeth$^2$, Wolfgang L\"{o}ffler$^1$ and Dirk Bouwmeester$^{1,2}$}
\affiliation{%
 $^1$Huygens-Kamerlingh Onnes Laboratorium, Leiden University, 2333 CA Leiden, The Netherlands.\\
 $^2$Department of Physics, University of California, Santa Barbara, CA 93106, USA.
}%






\maketitle

\renewcommand{\thefigure}{S\arabic{figure}}

\begin{figure*}[ht]
\includegraphics[scale=0.65]{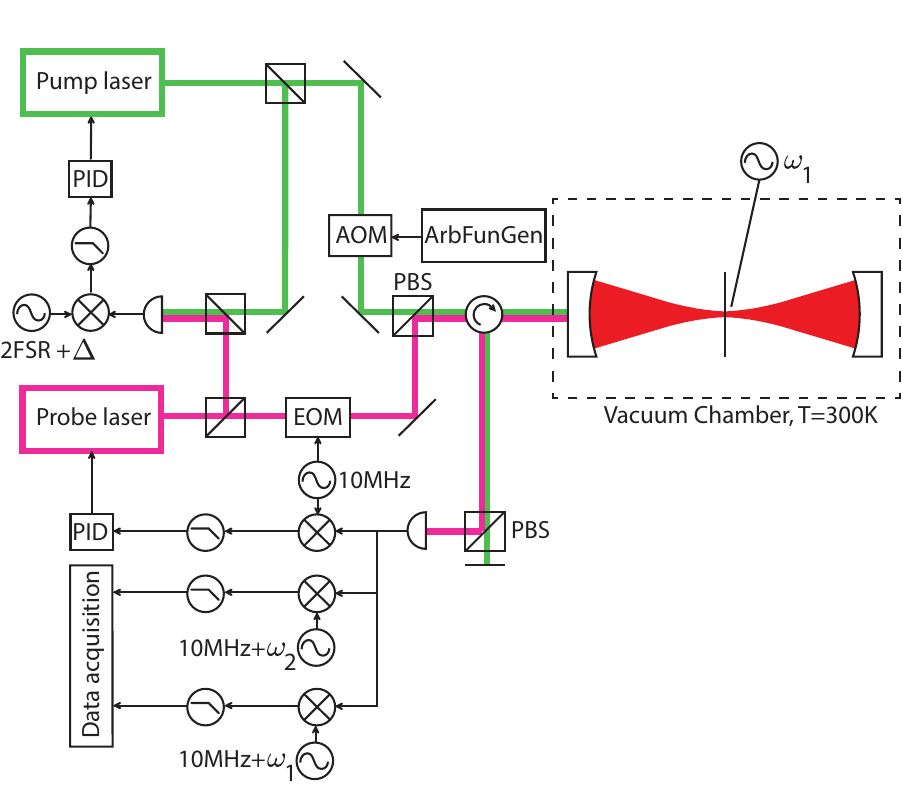}
\caption{Optical setup. The probe laser is locked to the transmission peak of the cavity. The pump laser is locked to the probe laser with frequency difference $\omega_{\mathrm{probe}}-\omega_{\mathrm{pump}}=2\mathrm{FSR}+\Delta$, controlled by an RF source. The driving pulses are shaped by electronic pulses sent to the AOM from an arbitrary wave function generator (ArbFunGen). Polarizing beam splitters (PBS) are used to separate the probe and pump light fields. Mechanical modes are excited by a needle placed close to the membrane defect.}
\label{fig_S1}
\end{figure*}

\section{Setup}

The motion of the membrane is read out via the light fields generated by a probe laser at $\omega_{\mathrm{probe}}$ (10 $\mathrm{\mu W}$) locked to the transmission resonance frequency of the optical cavity via the Pound-Drever-Hall technique (PDH) \cite{Drever1983}. In order to measure the instantaneous oscillation displacement of a mechanical mode $\omega$, the reflection signal is demodulated at $\omega + \omega_{\mathrm{EOM}}$, where $\omega_{\mathrm{EOM}}=10$ $\mathrm{MHz}$ is the detuning frequency of the sidebands in the PDH locking scheme. The driving light fields generated by the pump laser at $\omega_{\mathrm{pump}}$ is locked to the probe laser by a phase-locked loop, see Fig. \ref{fig_S1}. The light fields from the two lasers are measured by a fast photodetector and the beating signal is mixed with a reference microwave signal, supplied by an RF generator. The resulting signal is sent to a proportional-integral-differential controller (PID) which adjusts the frequency of the pump laser. The difference between the lasers frequencies is kept at $\omega_{\mathrm{probe}}-\omega_{\mathrm{pump}}=2\mathrm{\mathrm{FSR}}+\Delta \sim 3$ $\mathrm{GHz}$ because the dispersion curves of membrane-in-the-middle systems are parallel for all odd and all even resonances \cite{Jayich2008}. This ensures a well-defined cavity resonance detuning of the pump laser in spite of drifts in the membrane position along the optical axis ($\sim 10$ $\mathrm{nm/hour}$). The pump light fields have orthogonal polarization to the probe light fields in order to minimize interference of both fields at the reflection photodetector. To excite a membrane mechanical mode, an AC voltage ($\sim 10$ $\mathrm{mV}$) at its mechanical frequency is applied to a needle placed close to the defect of the membrane ($\sim 0.5$ mm). The full membrane 1,1 mechanical mode thermal motion is damped by applying an electrostatic force through the needle. The force is proportional to the instantaneous position of this mode but delayed quarter of its oscillation, which effectively creates a frictional force proportional the the mode's velocity.
 
\section{Membrane positioning}
The dispersion curves of a membrane-in-the-middle system are parallel for the curves separated by 2FSR provided the membrane is positioned exactly in the middle of the cavity. For a small displacement $z$ of the membrane from the center, the free-spectral range changes as $2\mathrm{FSR}-2\mathrm{FSR}_{\mathrm{middle}}\propto 2\mathrm{FSR}_{\mathrm{middle}}\frac{z}{L}\sin{\pi \frac{z}{\lambda}}$, where $L$ is the length of the cavity and $\lambda$ is the wavelength. The membrane holder is mounted on a tip-tilt stage with 3 vacuum compatible motors (1 step $\sim 20$ $\mathrm{nm}$). To minimize the influence of the membrane drift along the optical axis, the membrane was moved towards the middle of the cavity to $z\sim30$ $\mathrm{\mu m}$ by measuring 2FSR as a function of $z$, which provides an estimate for the direction and amplitude of the movement. To further minimize the influence of membrane drifts, we use a piezo element to bring the membrane to the position where 2FSR has a local maximum as a function $z$. This position coincides with the maximum optomechanical coupling strength. As a result an average drift of the membrane during a measurement run of 1 hour causes an acceptable change of 2FSR $\sim 5$ $\mathrm{kHz}$. When the actual experiment is running, we use the piezo to bring the membrane back to the position of maximum 2FSR every hour.

\section{Membrane fabrication}
We begin the fabrication process of the devices with a commercially supplied $525$ $\mathrm{\mu m}$ thick silicon wafer coated on both sides with $25$ $\mathrm{nm}$ of LPCVD high-stress silicon nitride. We pattern the phononic crystal structure into the nitride on one side through the use of standard photolithography. During a second photolithography step, we use an IR contact aligner to pattern a square hole in photoresist on the opposite side of the chip. A subsequent Bosch etch step etches through the exposed nitride and removes about $425$ $\mathrm{\mu m}$ of the silicon underneath the phononic crystal. After cleaning the chip in piranha solution, we release the phononic crystal membrane by wet etching the remaining $100$ $\mathrm{\mu m}$ of silicon using KOH at $80^\circ$C. We perform a final clean by submerging the chip in HF for 1 minute and then we extract it out of IPA and allow it to dry through evaporation.

\section{Detailed theory}
Here we derive Eq. (3) of the main text and the full model which accounts for the unmatched sidebands. We start from the optomechanical equations of motion \cite{Aspelmeyer2014} in the presence of two mechanical modes $\hat{b}_i$ and two coherent driving fields at $\omega_{\mathrm{L}1}$ and $\omega_{\mathrm{L}2}$ with the condition $\omega_{\mathrm{L}1}+\omega_1=\omega_{\mathrm{L}2}+\omega_2=\omega_{\mathrm{cav}}$, where $\omega_i$ is the frequency of mechanical mode $i$, $i$=1,2. In the linearized approximation and in the frame rotating at $\omega_{\mathrm{cav}}$, the total intracavity light fields $\hat a$ is
\begin{equation}
\hat{a}=|\bar\alpha_1(t)|e^{i(\omega_1t+\phi_1)}+|\bar\alpha_2(t)|e^{i(\omega_2t+\phi_2)}+\delta \hat{a},
\end{equation}
where $\bar\alpha_i$ is the amplitude of the intracavity field due to driving field $i$, $\phi_i$ is a constant and $\delta\hat{a}$ is a fluctuating term. The evolution of $\delta\hat a$ is given by
\begin{equation}
\delta\dot{\hat{a}}=-\frac{\kappa}{2}\delta\hat{a} + i(G_1\hat{x}_1+G_2\hat{x}_2) \hat{a},
\end{equation}
where $G_i$ is the optical frequency shift per displacement of the mechanical mode $\hat{x}_i=x_{zpm,i}(\hat b_i+\hat b^\dagger_i)$ with  $x_{zpm}$ being the zero-point motion of mode $i$. Neglecting the thermal occupation of the environment, the mechanical modes evolve as
\begin{equation}
\dot{\hat{b}}_i=(-\frac{\Gamma_i}{2}-i\omega_i)\hat{b}_i + ig_{0i} \hat{a}^\dagger\hat{a},
\end{equation}
where $g_{0i}$ is the single photon optomechanical coupling of mode $i$. In the frame rotating at $\omega_i$ for both mechanical modes  $\hat{c}_i=e^{i(\omega_it+\phi_i)}\hat b_i$
\begin{equation}
\dot{\hat{c}}_i=-\frac{\Gamma_i}{2}\hat{c}_i + ig_{0i} \hat{a}^\dagger\hat{a} e^{i(\omega_it+\phi_i)}.
\end{equation}
The sidebands at $\omega_{\mathrm{cav}}$ have much larger amplitude than the other sidebands. Thus RWA is a good approximation for this situation. Applying RWA and linearizing, we obtain
\begin{align*}
\hat{a}^\dagger\hat{a} e^{i(\omega_it+\phi_i)}&=\\
&=(|\bar\alpha_1|e^{-i(\omega_1t+\phi_1)}+|\bar\alpha_2|e^{-i(\omega_2t+\phi_2)}+\delta\hat{a}^\dagger)\times\\
&\times(|\bar\alpha_1|e^{i(\omega_1t+\phi_1)}+|\bar\alpha_2(t)|e^{i(\omega_2t+\phi_2)}+\delta \hat{a})\times\\
&\times e^{i(\omega_it+\phi_i)}=\\
&=|\bar{\alpha}_i|\delta\hat a
\end{align*}
\begin{align}
&\dot{\hat{c}}_i=-\frac{\Gamma_i}{2}\hat{c}_i + ig_{0i}|\bar{\alpha}_i|\delta\hat a\\
&\delta\dot{\hat{a}}=-\frac{\kappa}{2}\delta\hat{a} + iG_1 x_{zpm,1} |\bar{\alpha}_1|\hat c_1 + iG_2 x_{zpm,2} |\bar{\alpha}_2|\hat c_2
\end{align}
Using the multiphoton optomechanical coupling $g_i(t)=G_i x_{zpm,i} |\bar{\alpha}_i(t)|$, changing notation $\hat c\rightarrow \hat b$ and $\delta\hat a \rightarrow -\delta\hat a$ we get Eq. (3) of the main text:
\begin{align}
&i\dot{\hat{b}}_i=-i\frac{\Gamma_i}{2}\hat{b}_i + g_i(t)\delta\hat a\\
&i\delta\dot{\hat{a}}=-i\frac{\kappa}{2}\delta\hat{a} + g_1(t) \hat b_1 + g_2(t)\hat b_2.
\end{align}

FULL Model. The full model includes the unmatched sidebands by not using RWA. We start with the linearized equation for the light fields Eq. (6) and the expression $\hat{x}_i=x_{zpm,i}(\hat b_i+\hat b^\dagger_i)$. Using Eq. (5) we get

\begin{equation}
\begin{aligned}
\delta\dot{\hat{a}}&=-\frac{\kappa}{2}\delta\hat{a} + i(g_{01}(\hat{b}_1+\hat{b}_1^\dagger)+g_{02}(\hat{b}_2+\hat{b}_2^\dagger))\times\\
&\times(|\bar\alpha_1(t)|e^{i(\omega_1t+\phi_1)}+|\bar\alpha_2(t)|e^{i(\omega_2t+\phi_2)}),
\end{aligned}
\end{equation}
The dynamics of the mechanical modes is still described by Eq. (7). As before we change the frame by applying the transformation $\hat{c}_i=e^{i(\omega_it+\phi_i)}\hat b_i$. In the expansion of its last term $i g_{0i}\hat{a}^\dagger\hat{a}$ the terms not including $\delta\hat{a}$ can be omitted and we get
\begin{equation}
\begin{aligned}
\dot{\hat{c}}_i&=-\frac{\Gamma_i}{2}\hat{c}_i + i g_{0i}e^{i(\omega_it+\phi_i)}\times\\
&\times(|\bar\alpha_1(t)|e^{-i(\omega_1t+\phi_1)}\delta\hat{a}+|\bar\alpha_2(t)|e^{-i(\omega_2t+\phi_2)}\delta\hat{a}+\\
&+|\bar\alpha_1(t)|e^{i(\omega_1t+\phi_1)}\delta\hat{a}^\dagger+|\bar\alpha_2(t)|e^{i(\omega_2t+\phi_2)}\delta\hat{a}^\dagger)
\end{aligned}
\end{equation}
To simulate this model we average the operators $\hat{a}$, $\hat{c}$ to get the classical fields. Simulation of this model for the experimental parameters is done by solving these differential equations. 

Transfer efficiencies calculated by the full model start to deviate from ones by Eq. (3) of the main text by more than 3\% for pulses with $\sigma\gtrsim 25$msec. We see negligibly small dependence of transfer efficiency on $\phi_1-\phi_2$.

Note, simulations show that using Eq. (3) of the main text with added corrections due to the optical spring effect of the unmatched sidebands give incorrect result.


\section{Quantum simulations of STIRAP of single-phonon Fock state}
In the following, we provide a full quantum simulation of STIRAP for a Fock-like state of a mechanical mode. We take into account known experimental complications including: thermalization to the environment, heating of the membrane by driving fields, presence of other mechanical modes in the membrane, realistic detection efficiency of the heralding photons, and dark count rate (DCR) of the detectors.

We assume that the system is cryogenically cooled to 20mK and that the membrane is heated, conservatively, to 1K by the pump and probe driving fields \cite{Laser_Cooling} \cite{Measurement_Motional}. For the optomechanical parameters considered in this work, the average phonon occupation of mechanical modes at equilibrium with the environment is much larger than unity, even at mK temperatures. Thus, thermalization to the environment (leakage of phonons from the environment to mechanical modes) must be taken into account to realize the state transfer of a mechanical Fock-like state. This effect requires us to further optically cool the mechanical modes of interest.

To perform STIRAP in the quantum regime, we must develop a full experimental procedure for state transfer. This procedure would contain three steps: State preparation, STIRAP, and state readout. We consider, in the following, two defect modes of a phononic crystal membrane at $\omega_1/2\pi=1.2$ MHz and $\omega_2/2\pi=1.4$ MHz. These modes have quality factors $Q_1 = Q_2 = 10^9$ \cite{Measurement-based} and are in a thermal bath at $T=1$K. We consider a cavity with linewidth, $\kappa/2\pi=50$ kHz, and optomechanical single-photon coupling strengths, $g_{01}/ 2\pi=g_{02}/ 2\pi=1$ Hz.

{\bf State Preperation:} We begin by cooling our system, comprised of a single phononic crystal membrane and optical cavity, using a red detuned laser ($\Delta = -\omega_m$) in the resolved sideband regime ($\kappa<<\omega_m$). This creates an interaction between the cavity and resonator, given by the linearized interaction Hamiltonian \cite{Aspelmeyer2014}:

\begin{equation}
    H_{int} \propto \hat{a}^{\dagger} \hat{b}+\hat{a} \hat{b}^{\dagger}.
\end{equation}
Here $\hat{a}$ ($\hat{a}^{\dagger}$) represents the annihilation (creation) operator for the cavity mode and $\hat{b}$ ($\hat{b}^{\dagger}$) represents the annihilation (creation) operator for a mechanical defect mode of the membrane.
In the sideband resolved regime the ratio of the rates of anti-Stokes (blue-shifted) to Stokes (red-shifted) photons created by this interaction tends towards infinity, as the rate of Stokes photons created in the cavity approaches zero and the rate of anti-Stokes photons approaches the flux of thermal phonons from the environment \cite{Polzik}. This sets an optical cooling limit, for our system in the sideband resolved regime, given by the steady-state final phonon population  \cite{Laser_Cooling} \cite{Aspelmeyer2014}:
\begin{equation}
    \Bar{n}_{f} = \frac{\Gamma_{opt} \Bar{n}_{min}+ \Bar{n}_{th}\Gamma_{m}}{\Gamma_{opt}+\Gamma_{m}}.
\end{equation}
Here $\Gamma_m$ is the mechanical loss of the defect mode, $ \Bar{n}_{th}$ is the average thermal occupation of our environment, and $\Gamma_{opt}$ and $\Bar{n}_{min}$ are the minimum phonon number of a mechanical mode and optomechanical damping rate respectively defined as \cite{Aspelmeyer2014},

\begin{equation}
\begin{split}
    \Gamma_{opt} &= g^2\Big{(}\frac{\kappa}{\kappa^2/4+(\Delta+\omega_m)^2}-\frac{\kappa}{\kappa^2/4+(\Delta-\omega_m)^2}\Big{)}\\
    \Bar{n}_{min} &= \Big{(}\frac{\kappa^2/4+(\Delta-\omega_m)^2}{\kappa^2/4+(\Delta+\omega_m)^2}-1\Big{)}^{-1}
\end{split}
\end{equation}
where $g=\alpha g_0$ is the light-enhanced optomechanical coupling (in terms of $g_0$, the optomechanical single-photon coupling strength, and $\alpha$, the maximum laser drive intensity), $\kappa$ is the cavity intensity decay rate, $\Delta$ is the laser detuning from resonance (taken to be $\pm\omega_m$ in our case), and $\omega_m$ is the frequency of the defect mode.

Thermal steady states of average occupation, $\Bar{n}^{r}_{f} \approx 0.10$ \cite{Laser_Cooling} \cite{Polzik} \cite{ground_state_theory} have been shown to be experimentally achievable. For our system, steady state occupation can be demonstrated with a 5 ms red-detuned ($\Delta = -\omega_1$) constant pulse of $g^{r} /2\pi= 1600$ Hz, which is realistic if the frequency noise of the  pump light is reduced by a narrow bandwidth filtering cavity. This corresponds to a density matrix of the form,

\begin{equation}
\label{red}
\begin{split}
    \rho_{1}^{r} =  0.906\ket{0}\bra{0} + 0.085\ket{1}\bra{1} + 0.008\ket{2}\bra{2} + \ldots \\
    \rho_{2}^{r} =  0.906\ket{0}\bra{0} + 0.085\ket{1}\bra{1} + 0.008\ket{2}\bra{2} + \ldots \\
\end{split}
\end{equation}
Where $\rho_{1}^{r}$ corresponds to a defect mode at $\omega_1$ and $\rho_{2}^{r}$ corresponds to a defect mode at $\omega_2$. It is clear from $\rho^{r}$ that the single phonon occupation is much larger than the two phonon occupation.

Next this state is converted to a single phonon Fock-like state. Detection of a heralded Stokes photon, while pumping with strong red-detuned driving fields, would project this system to the desired state \cite{Heralded_prep}, but the flux of the Stokes photons is negligible as noted above. A short constant blue-detuned ($\Delta = \omega_1$, $\tau_b\approx$ 0.1ms and $g^{b}/2\pi=2410$ Hz) pulse raises the occupation of our state to, $n_{f1} \approx$ 0.20 and $n_{f2} \approx$ 0.12. This corresponds to a probability of ~0.1 for a Stokes photon to be created  by mode 1 (the probability of two Stokes photons is ~$0.1^2$). After the pulse mode 1 and 2 density matrices are given by $\rho_{1}^b$ and $\rho_{2}^b$,
\begin{equation}
\label{bluenum}
\begin{split}
    \rho_{1}^{b} = 0.831\ket{0}\bra{0} + 0.140\ket{1}\bra{1} + 0.024\ket{2}\bra{2} + \ldots \\
    \rho_{2}^{b} = 0.893\ket{0}\bra{0} + 0.095\ket{1}\bra{1} + 0.010\ket{2}\bra{2} + \ldots
\end{split}
\end{equation}



which includes thermalization from the environment, described below.

Once a projective measurement has been made by a single photon detector (SPD) on our state, we can collapse the vacuum amplitude, as we deterministically know that a phonon has been created by the Stokes scattering process in mode 1.
This will have a high probability of being a single phonon, as the two phonon probability amplitude is much less than the single phonon probability amplitude seen in Eq. (\ref{bluenum}) \cite{Remote}. This configured density matrix would correspond to,
\begin{equation}
\label{conf}
    \rho_{1}^{conf} = 0.831\ket{1}\bra{1} + 0.140\ket{2}\bra{2} + 0.024\ket{3}\bra{3} + \ldots
\end{equation}

The generated Stokes photon at $\omega_{cav}$ is accompanied by strong pump fields approximately $10^9$ stronger due to low scattering probability ~$(g_0/ \kappa)^2$. To filter out the pump light fields, the light transmitted through the cavity with the membrane is sent through a set of consecutive narrow-linewidth filtering cavities as demonstrated in \cite{Polzik}, where 150dB suppression of light at 1.5MHz was shown.

The DCR of the detector and overall detection efficiency during heralding must be considered to provide a realistic estimate to whether or not experimentation is achievable. The overall detection efficiency ($\eta$) is set by the detection efficiency of the single photon detector and the optics of our system including the filtering cavities. Similar experimentation has achieved an overall detection efficiency of 2.5\% \cite{Polzik}, however this can be increased to 7.5\% by using superconducting nanowire SPD's. This detection efficiency of 7.5\% will be used throughout the following.
If the detector clicks due to dark counts, the state of the first mode is unchanged from $\rho_b^1$. The rate of detected Stokes photons from the heralding is 75Hz ($(\tau_{b}/p\eta)^{-1}$), where $p\approx0.1$ is the probability that a single Stokes photon is created). Taking into account a DCR of 10 Hz we would see an overall configured state of,

\begin{equation}
\begin{split}
      \rho_1^{i}&= \frac{10\mathrm{Hz}}{10\mathrm{Hz}+75\mathrm{Hz}}\rho_1^b +\frac{75\mathrm{Hz}}{10\mathrm{Hz}+75\mathrm{Hz}}\rho_1^{conf}\\
    \rho_1^{i}&=0.098\ket{0}\bra{0} + 0.750\ket{1}\bra{1} + 0.126\ket{2}\bra{2} + \ldots  
\end{split}
\end{equation}

{\bf STIRAP:} In order to realize STIRAP numerically we make use of the Quantum Toolbox in Python (QuTiP) \cite{QuTiP}.
Using annihilation and creations operators for the mechanical defect modes and optical cavity, we write a linearized optomechanical interaction Hamiltonian, defined in a frame rotating w.r.t the mechanical modes and cavity (triply rotating frame) \cite{multimode}. This Hamiltonian can be written as:
\begin{equation}
\label{parametric}
H_{s} = \sum_{i,j=1}^2g_{0j}\alpha_{i}(\hat{a}^{\dagger} e^{i{\Delta_i}t} + \hat{a} e^{-i{\Delta_i}t} )(\hat{b}_j e^{-i\omega_{j}t} + \hat{b}_{j}^{\dagger} e^{i\omega_{j}t})
\end{equation}
where,
\begin{equation}
\label{fSTIRAP pump parameters}
\begin{split}
\alpha_1(t) &= \alpha_{max} e^{{-(t - \tau)}^2/2\sigma^2}\\
\alpha_2(t) &= \alpha_{max} e^{{-(t + \tau)}^2/2\sigma^2}
\end{split}
\end{equation}
are the Gaussian pumps used to realize STIRAP. Here $\sigma=$0.14ms, $\tau=\sigma/3$, $\alpha_{max}=5000$.

QuTiP provides several ways to solve quantum dynamics, however, as we will eventually examine degrees of entanglement in our system, it is convenient to work in density matrix formalism, which necessitates solving a master equation. Specifically, we numerically integrate a Lindblad master equation of the form \cite{Heralded_prep},
\begin{equation}
\label{master}
\dot{\rho}(t) = \frac{-i}{\hbar}\left[H_s,\rho\right] + \frac{\kappa}{2} \mathcal{L}(\hat{a}) + \sum_{i=1}^2 \frac{\Gamma_{m_i}}{2}\Big{(}(\bar{n}_i + 1) \mathcal{L}(\hat{b}_i) + \bar{n}_i\mathcal{L}(\hat{b}_i^\dagger)\Big{)}
\end{equation}
where $\mathcal{L}(\hat{O}) = \frac{1}{2} (2 \hat{O} \rho(t) {\hat{O}}^{\dagger} - \rho(t) {\hat{O}}^\dagger \hat{O} - {\hat{O}}^\dagger \hat{O} \rho(t))$ are superoperators and $\bar{n}_i = ({{e^{{h\omega_i}/{k_bT}}-1}})^{-1}$ is the average number of thermal phonons at the ith mechanical frequency.

\begin{figure*}[!hbt]
\centering
\begin{minipage}[bt!]{8cm}
\centering
\includegraphics[width=8cm]{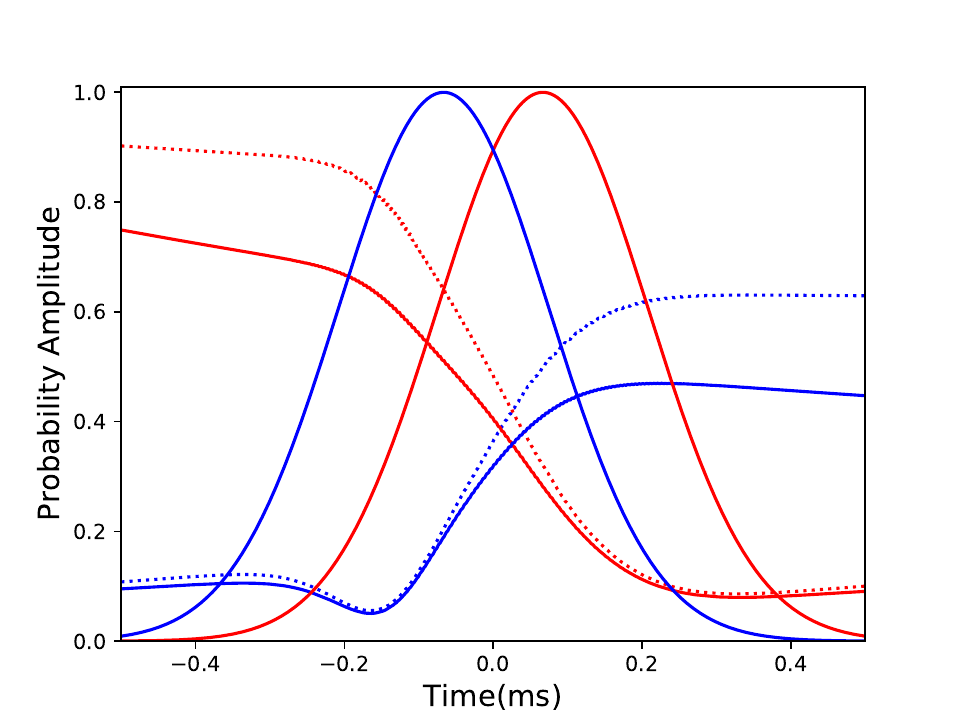}
\caption{State transfer using STIRAP at 1K. Red and blue represent the first and second modes respectively. Solid denotes the probability of detecting a single phonon, while dotted denotes the probability that the mode is populated ($1-\rho_{00}$). The efficiency of state transfer is 59.6\% and includes unmatched sidebands.}
\label{STIRAP_realistic_1K_nRWA}
\end{minipage}\hfill
\begin{minipage}[bt!]{8cm}
\centering
\includegraphics[width=8cm]{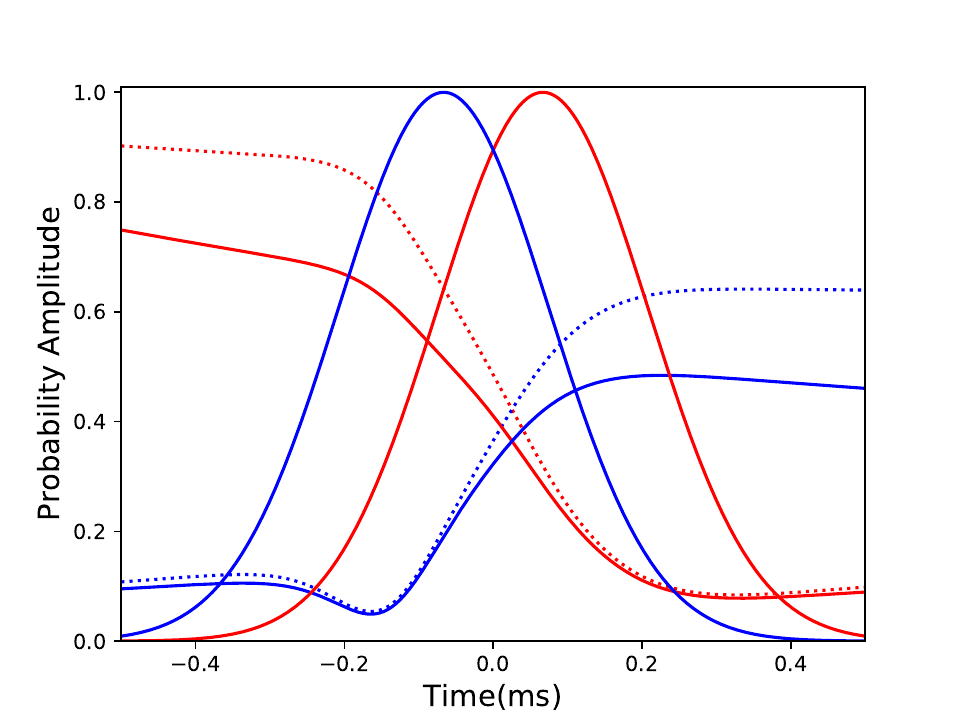}
\caption{State transfer using STIRAP at 1K and the RWA approximation. Red and blue represent the first and second modes respectively. Solid denotes the probability of detecting a single phonon, while dotted denotes the probability that the mode is populated ($1-\rho_{00}$). The efficiency of state transfer is 60.1\%.}
\label{STIRAP_realistic_1K_RWA}
\end{minipage}
\end{figure*}

We simulate Eq. (\ref{master}) with an initial state of $\rho_1^i$ and $\rho_2^b$, and produce Fig. \ref{STIRAP_realistic_1K_nRWA}, which shows the transfer process. The single phonon amplitude transfer has an efficiency of 59.6\%, which is predominately due to thermalisation and low pumping power (limited by the cavity linewidth). These two factors must be balanced as: If the resonator is pumped too hard we introduce thermal phonons into the mechanical mode, if we pump too long thermalization will introduce phonons from the environment, but if we pump harder or longer we will see a higher degree of state transfer.
With these factors accounted for the final density matrices are,

\begin{equation}
\begin{split}
    \rho_1^{s}= 0.900\ket{0}\bra{0} + 0.091\ket{1}\bra{1} + 0.007\ket{2}\bra{2} + \ldots \\ \\
    \rho_2^{s}= 0.370\ket{0}\bra{0} + 0.447\ket{1}\bra{1} + 0.137\ket{2}\bra{2} + \ldots
\end{split}
\end{equation}

The case outlined here is conservative in the parameters that we have chosen, and we would expect to achieve similar, if not better, results when performing this experiment in the future.

Additionally, we can apply the rotating wave approximation (RWA) to Eq. (\ref{parametric}) to confirm whether or not the unmatched sidebands negligibly effect our system. This approximation produces,
\begin{equation}
\label{RWA}
H_{s}^{rwa} = \sum_{i,j=1}^2g_{j}\alpha_{i}(\hat{a}^\dagger \hat{b}_j + \hat{a} \hat{b}^\dagger_j).
\end{equation}
which is simulated using Eq. (\ref{master}). This process has a single phonon amplitude transfer efficiency of 60.1\%, see Fig. \ref{STIRAP_realistic_1K_RWA}, which is a negligible increase when compared to when the unmatched sidebands are included (Fig. \ref{STIRAP_realistic_1K_nRWA}).

The Lindblad master equation requires that the secular approximation (RWA) is applicable to the system being examined \cite{Secular}. For our case this is shown to be true, however (comparing the cases when the RWA is applied and when the unmatched sidebands are included) if the modes in the phononic bandgap were closer together ($\kappa \approx \omega_2 -\omega_1$), this approximation would fail. This emphasises the importance that the modes be far enough apart to apply the secular approximation, but not so far that we transfer population to nearby modes outside the phononic bandgap.

\begin{figure*}[!hbt]
\centering
\begin{minipage}[bt!]{8cm}
\centering
\includegraphics[width=8cm]{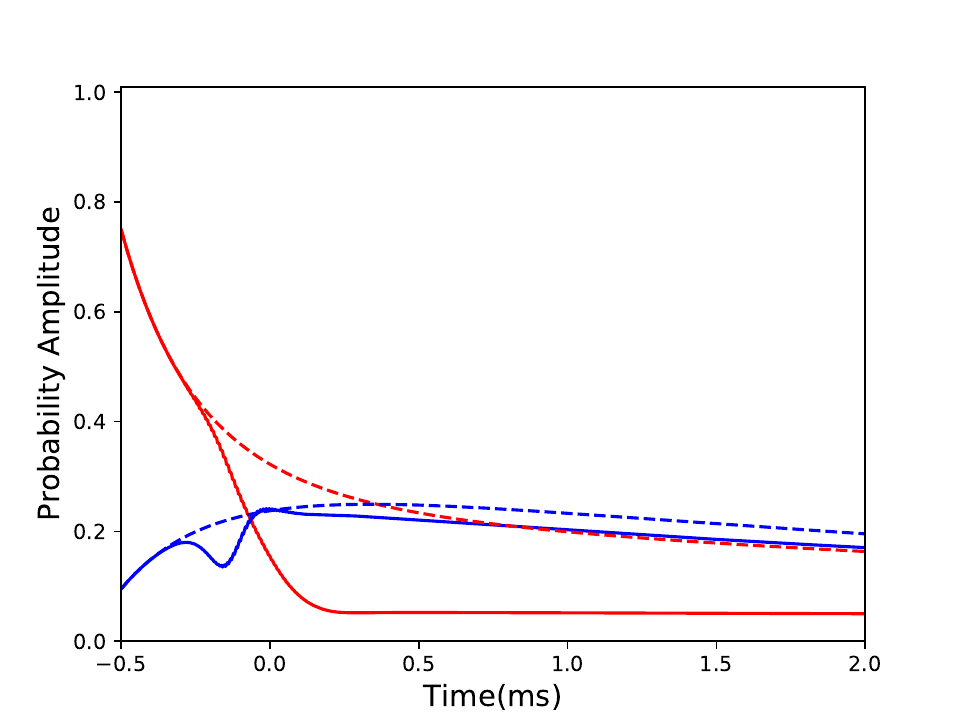}
\caption{State transfer at 8K realized using STIRAP. Red and blue represent the first and second modes respectively. Solid denotes a state transfer via STIRAP, while dashed denotes evolution in the presence of the environment, without the STIRAP pulses.}
\label{STIRAP_realistic_8K_nRWA}
\end{minipage}\hfill
\begin{minipage}[bt!]{8cm}
\centering
\includegraphics[width=8cm]{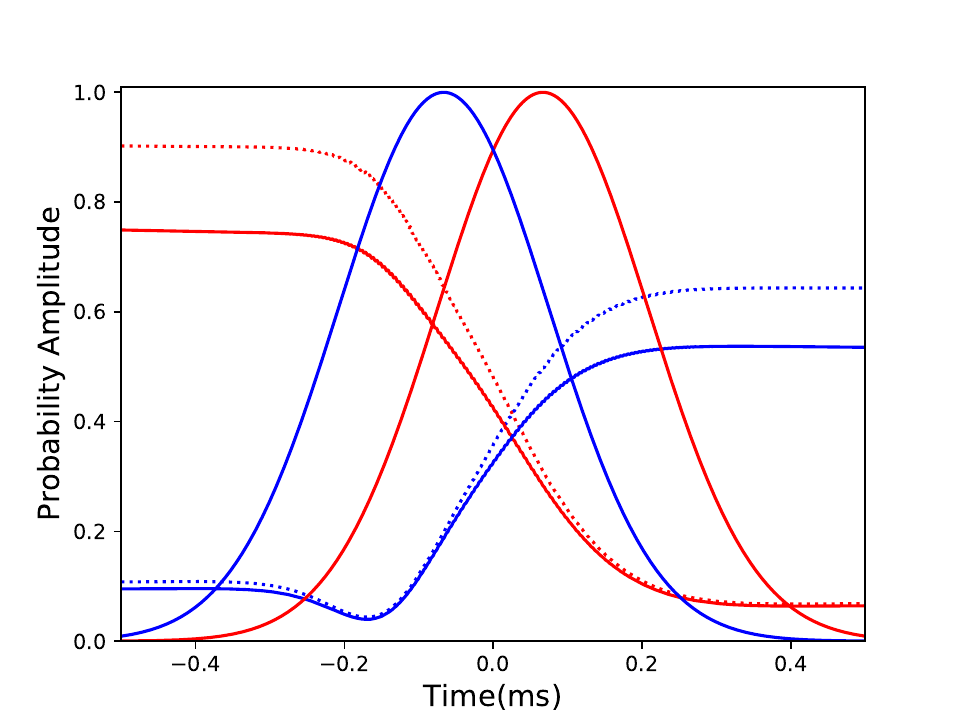}
\caption{State transfer using STIRAP at 0.1K. Red and blue represent the first and second modes respectively. Solid denotes the probability of detecting a single phonon, while dotted denotes the probability that the mode is populated ($1-\rho_{00}$). The efficiency of state transfer is 71.3\% and includes unmatched sidebands, which contributes the oscillation seen in either amplitude.}
\label{STIRAP_realistic_0.1K_nRWA}
\end{minipage}\hfill
\end{figure*}

We determine the behavior of our system at 8K to emphasize the need for state of the art cryogenic cooling and low driving power of the mechanical resonator. Due to the thermal environment, we would expect to see STIRAP break down and thermalize at a rate determined by our temperature and with an average occupancy determined by the mechanical dimensionality of our numeric integration (35). This produces Fig. \ref{STIRAP_realistic_8K_nRWA}, which shows a comparison between STIRAP and our system's evolution without the presence of the STIRAP pulses. At this temperature it would be difficult (if not impossible) to determine any deviation between these two cases experimentally, and sets a limitation to our mechanical resonator's temperature at $\approx$ 3K, using the realistic parameters we have detailed.

Finally, we demonstrate an optimistic experiment where the bath temperature is 100 mK. This produces Fig. \ref{STIRAP_realistic_0.1K_nRWA}, which has a single phonon state transfer efficiency of 71.3\%. Additionally, it can be seen that all other occupation is low in comparison to Fig. \ref{STIRAP_realistic_1K_nRWA}. This improves the quality of experimentation, as the two phonon and greater populations would have a reduced effect in detection. By comparing these three cases, we can determine that the most important parameter, in the context of transfer efficiency once the adiabatic parameters have been satisfied, is the temperature of the environment.

{\bf State Readout:} In order to readout the transferred state, we employ a short red detuned pulse ($\approx 0.5$ms, $g_1 / 2\pi=5000$Hz) to transfer the single phonon population into anti-Stokes photons with probability 99.8\% (calculated from $e^{-\Gamma_{opt}\tau}$). This pulse duration is chosen to optimize the balance between heating effects $(\Gamma_m*n_{th}*0.5\mathrm{ms} \approx0.06$ phonons will enter into a mode by thermalization) and  the probability of converting phonon population into anti-Stokes photons. The anti-Stokes photons created through this process, would then be fiber collected and detected by super conducting SPDs (SSPD). As the detector cannot resolve the photon numbers, we show the dotted lines in Fig. \ref{STIRAP_realistic_1K_nRWA}, \ref{STIRAP_realistic_1K_RWA} and \ref{STIRAP_realistic_0.1K_nRWA}, showing the probability of creation of any non-zero number of Stokes photons ($1-\rho_{00}$). The rate of the detector clicking during the readout is proportional to this probability (if DCR is not accounted). 


The DCR is a main feature of SSPD's that usually presents a challenge in low (Mhz) frequency experimentation \cite{interferometry}. Conservatively, we would expect the DCR to be ~10 Hz for our system \cite{Superconducting_DCR} \cite{Ultimate_DCR}. On the time scale of our experiment during heralding (0.1 ms) and readout (0.5 ms), we would expect to see the DCR in combination with 75 Hz (heralding) and 15 Hz (readout) signals. These two signal rates are found by combining the detection efficiency with the two timescales, as shown above.

The DCR is non-negligible in comparison to the signal rates. However, in the case of the state preparation heralding the DCR is taken into account and as the DCR would be independently measured, and is constant, it can be subtracted from the readout signal rate. Additionally, it has been shown that by operating SSPD's with a low bias current, in combination with a cryogenically cooled optical band-pass filter, the DCR can be reduced to an almost negligible level ($10^{-4}$)\cite{Ultimate_DCR}.

Timing considerations are a major challenge for low frequency experimentation \cite{interferometry}. However, the time to get a single click (using our system parameters) from blue-detuned heralding can be estimated from the time of detection, efficiency, and probability of creating a Stokes photon. 

\begin{equation}
    T_{h}=\frac{5\mathrm{ms}+0.1\mathrm{ms}}{0.1\eta}=0.7\mathrm{s}
\end{equation}

Following a detection event of the Stokes photon, the STIRAP sequence and readout pulse are sent, which results in a final probability of detection of:

\begin{equation}
    P_f=\eta(1-\rho_{00})<0.075
\end{equation}

This sets a time for a click at readout of $T_r=T_h/P_f > 9.3$s depending on the density matrix to be measured. This time would allow us to measure the data needed to recreate the state transfer process depicted by the dotted lines in Fig. \ref{STIRAP_realistic_1K_nRWA}.

{\bf Other Membrane Modes.} Now we consider the presence of nearby mechanical modes in the membrane. The STIRAP protocol relies on the heralding Stokes photons from the blue-detuned pulse and anti-Stokes photons from the read-out pulse. These two pulses affect other modes of the membrane in the frequency range of a few $\kappa$ around $\omega_1$ and $\omega_2$. As a consequence, a flux of spurious Stokes and anti-Stokes photons is produced by these modes. We have observed modes in our membranes with quality factors as low as $10^6$ in the vicinity of mode 1 and 2. The upper bound of the rate for the phonons to enter these modes from the environment is ~100 kHz (based on $\Phi\approx\Gamma_m n_{th}$ where $n_{th}=10^5$). Additionally, the upper limit for the conversion of phonons into photons for these modes is $\Gamma_{opt} / 2\pi<$2 kHz.

Because we are operating in the bandgap of our membrane, the spurious modes are at least 70 kHz away from modes 1 and 2. This is also the minimal frequency separation of the spurious photons from $\omega_{cav}$, where the heralded photons are detected passing through the cascade of filtering cavities described above. These filtering cavities produce at least 50dB of isolation at 70 kHz \cite{Polzik}, which decreases the flux of the spurious photons $10^5$ times not taking into account the detection efficiency. This causes the rate of detector clicks due to other modes in the membrane to become negligibly small and does not effect the STIRAP protocol.

We have detailed a full protocol for implementing STIRAP in the quantum regime, taking into account all relevant experimental difficulties associated with low temperature experimentation we can foresee. In conclusion STIRAP of a phonon Fock-like state is feasible with state-of-the-art membranes available today. The major experimental challenge is keeping the temperature of the membrane at 1K or lower.

\section{Generation of driving pulses}
\begin{figure*}[ht]
\centering
\includegraphics[scale=1]{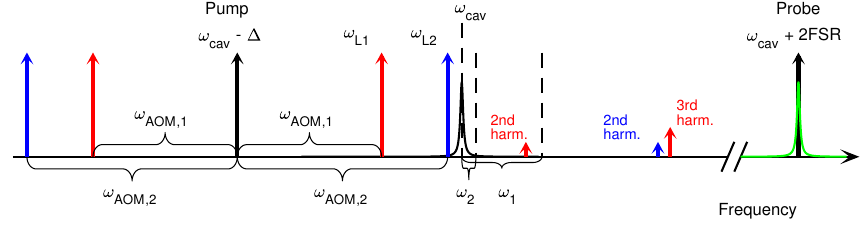}
\caption{Scheme of the optical frequencies. The probe laser is locked to the cavity resonance at $\omega_{\mathrm{cav}}+2$FSR. The amplitude of the pump laser at $\omega_{\mathrm{cav}}-\Delta$ is fully modulated by an AOM driven with AC voltage at $\omega_{\mathrm{AOM},i}$, thus only the light fields at $\omega_{\mathrm{pump}}\pm\omega_{\mathrm{AOM},i}$ reach the cavity. The upper sidebands ($\omega_{\mathrm{pump}}+\omega_{\mathrm{AOM},i}$) drive the state transfer, while the unwanted light fields at $\omega_{\mathrm{pump}}-\omega_{\mathrm{AOM},i}$ have a negligible effect due to their large detuning. The nonlinear response of the AOM leads to harmonics (small red and blue arrows) which we measure to also have a negligible effect.}
\label{fig_S6}
\end{figure*}

As mentioned in the main text, fluctuations in the difference of the frequencies of the two driving pulses must be much less than $1/T_{transfer}$ for the adiabaticity condition to be satisfied\cite{Vitanov2017}. We achieve this by generating both driving pulses from the same pump laser by frequency shift, see Fig. \ref{fig_S6}. An AC voltage with frequency $\omega_{\mathrm{AOM}}$ generates two light fields in the first diffraction maximum of the acousto-optical modulator (AOM) with frequencies $\omega_{\mathrm{pump}} \pm \omega_{\mathrm{AOM}}$. In order to independently address both frequencies required for the state transfer ($\omega_{\mathrm{L}1}$ and $\omega_{\mathrm{L}2}$), we send two electronic pulses to the AOM with frequencies $\omega_{\mathrm{AOM},i}$, $i=1,2$ and Gaussian envelopes generated by an arbitrary function generator (ArbFunGen). The pump laser detuning $\Delta=3.5$ $\mathrm{MHz}$ is chosen so that $\omega_{\mathrm{L}i}=\omega_{\mathrm{pump}}+\omega_{\mathrm{AOM},i}=\omega_{\mathrm{cav}}-\omega_i$ for mechanical modes at $\omega_i$, $i=1,2$, and the effect on the transfer process of the other pair of light fields at $\omega_{\mathrm{pump}} - \omega_{\mathrm{AOM},i}$ and harmonics $\omega_{\mathrm{pump}} + k\cdot\omega_{\mathrm{AOM},i}$, $k=2,3,4,...$ is negligible. The measured amplitude of the $2^{\mathrm{nd}}$ harmonics is much smaller than that of the $3^{\mathrm{rd}}$ harmonics, as is represented by the arrows labeled ``harm." in Fig. \ref{fig_S6}. To check the effect of the harmonics, we excite the mechanical modes to a level much higher than the thermal motion and we send driving pulses individually during the mechanical decay. With the above shown value for $\Delta$, mode 2 is not affected by the pulse sent to the AOM at $\omega_{\mathrm{AOM},1}$ within detection sensitivity, while the measured effect of the pulse at $\omega_{\mathrm{AOM},2}$ on mode 1 agrees well with the theoretically predicted optomechanical effect from the light fields at $\omega_{\mathrm{L}2}$.

\section{Calibration procedure}
\begin{figure*}[ht]
\centering
\includegraphics[scale=1]{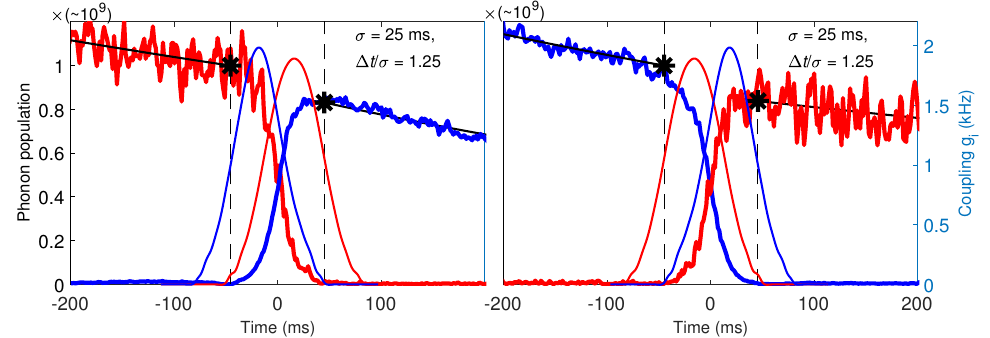}
\caption{Representative single runs of state transfer from mode 1 to mode 2 (left) and in the opposite direction (right). 
Left scale, thick lines: phonon population as a function of time, red line corresponds to mode 1, blue line to mode 2, both divided by the phonon population of mode 1 in the beginning of the transfer. Right scale, thin lines: multiphoton optomechanical couplings $g_1(t)$ red line, $g_2(t)$ blue line. The driving field pulses have a nearly Gaussian temporal profile, but their beginning and ending are modified such that they have zero amplitude outside the pulse. Vertical lines indicate the beginning and ending of the transfer process. Black stars correspond to the phonon populations used to calculate transfer efficiency.}
\label{fig_S7}
\end{figure*}

The transfer efficiency is defined in the main text as the ratio of the phonon population in mode 2 at the end of the transfer process to the phonon population in mode 1 at the beginning. The number of phonons in a mechanical mode $\langle \hat{b}_i^\dagger\hat{b}_i\rangle+\frac{1}{2}\propto u_i^2\propto R_i^2$, where $u_i$ is the amplitude of oscillation, and $R_i$ is the amplitude of the demodulated reflection signal measured at $\omega_i + \omega_{\mathrm{EOM}}$. Thus the transfer efficiency is 
\begin{equation}
\mathrm{Eff_{1\rightarrow2}}=\frac{k_2 R_2^2(t_{end,1\rightarrow2})}{k_1 R_1^2(t_{beginning,1\rightarrow2})},
\end{equation}
where the state is transferred from mode 1 to mode 2 and $k_i$ are coefficients of proportionality. Let us consider the reverse transfer $2\rightarrow1$. The product of transfer efficiencies
\begin{equation}
\mathrm{Eff_{1\rightarrow2}}\mathrm{Eff_{2\rightarrow1}}=\frac{R_2^2(t_{end,1\rightarrow2})}{R_1^2(t_{beginning,1\rightarrow2})}\frac{R_1^2(t_{end,2\rightarrow1})}{R_2^2(t_{beginning,2\rightarrow1})}
\end{equation}
does not have any coefficients of proportionality, thus it can be measured directly without any calibration. For the parameters of the transfer $\sigma=25$ $\mathrm{ms}$ and $\Delta t/\sigma=1.25$, this product is measured to be $0.73\pm0.05$. This implies that we demonstrate a transfer efficiency of at least $\sqrt{\mathrm{Eff_{1\rightarrow2}}\mathrm{Eff_{2\rightarrow1}}}=0.855\pm0.03$, independently of the model and calibrations. A numerical solution of the full model shows that  for the above chosen $\sigma$ and $\Delta t$, the efficiencies $\mathrm{Eff_{1\rightarrow2}}$ and $\mathrm{Eff_{2\rightarrow1}}$ differ by 0.01, which amounts to the transfer efficiency from the defect mode to the 3,3 mode being $0.86\pm0.03$, see Fig. \ref{fig_S7}.

The AOM used to shape the driving pulses has a non-linear intensity vs voltage response, which causes the actual temporal profile of the pulse's intensity to deviate from a Gaussian shape. Another consequence of this non-linearity is that the sum of intensities of individual pulses is not equal to the intensity of the pulse resulting from two Gaussian pulses being added and sent to the AOM. To account for these undesired effects, we measured the time profiles of the multiphoton optomechanical coupling $g_i(t)$ as follows. We excite mode 1 to a level much higher than the thermal occupation. During the mechanical decay, we send a single short Gaussian pulse $g_1(t, \sigma)$ to the cavity, with frequency $\omega_{\mathrm{L}1}$ and the same peak intensity as used for the STIRAP measurements. We adjust $\sigma$ for this pulse so that exactly half of the initial excitation energy is lost due to the optomechanical damping. This gives $\sigma_{1,1/2}=0.12\pm0.01$ $\mathrm{ms}$. Numerical solution of Eq. (3) of the main text for such a pulse gives the peak value of the pulse $\max{g_1(t)}\sim2$ $\mathrm{kHz}$. Next a similar procedure is followed for mode 2, but $\sigma$ of the pulse is set equal to $\sigma_{1,1/2}$, and the peak value of the pulse is set so that exactly half of the initial excitation of mode 2 is lost after the pulse $g_2(t, \sigma_{1,1/2})$ at $\omega_{\mathrm{L}2}$. This gives the estimate of $\max{g_2(t)}\sim 2$ $\mathrm{kHz}$ and the required voltage amplitude sent to AOM in the pulse. 

To get the actual temporal profile of $g_i(t)$, we measure in transmission the time profiles of the intensities of the pulses used for the transfer, with $\sigma=25$ $\mathrm{ms}$ and all the values of $\Delta t/\sigma$ used for the measurements (-1:0.25:4). In order to measure the exact temporal intensity profile of both STIRAP pulses individually, while both pulses are simultaneously applied (STIRAP sequence), the pump laser detuning $\Delta$ is adjusted such that $\omega_{\mathrm{cav}}-\omega_{\mathrm{L}1}\sim \kappa$, while $\omega_{\mathrm{L}2}+\omega_2=\omega_{\mathrm{L}1}+\omega_1$ as always, making $|\omega_{\mathrm{cav}}-\omega_{\mathrm{L}2}|\gg \kappa$. Therefore the transmitted light consists almost exclusively of the intensity at $\omega_{\mathrm{L}1}$. To correct for the small fraction of light at $\omega_{\mathrm{L}2}$, we send this pulse individually with the same detunings, and subtract the measured transmission from the case when both pulses are present. We follow the same procedure in order to measure the individual intensity of light at $\omega_{\mathrm{L}2}$. The measured intensity profiles of the pulses are used in the numerical simulations presented here and in the main text.

\section{Membrane heating and non-linear effects}
We observe increasing discrepancy of measured and simulated data for the state transfer with $\sigma\gtrsim 25$msec. This is caused by membrane heating by the driving pulses and by the defect mode frequency dependence on the amplitude of the full membrane 3,3 mode. Driving pulses heat the area of the membrane in the vicinity of the defect which decreases its frequency by $\sim 5$Hz (out of 1.25MHz) through thermal expansion leading to a decrease in the local stress. We observed a frequency change of the defect mode persisting for some time after a driving pulse. The other effect that changes the frequency of the defect mode is non-linearity of the membrane. In the realization of STIRAP, the level of excitation of the 3,3 mode should be much higher than its thermal occupation. This requires relatively large amplitudes of the 3,3 mode which effects the frequency of the defect mode through increased stress in the membrane averaged over an oscillation of the 3,3 mode. We observed an increase of the frequency of the defect mode when the 3,3 mode is excited by a couple of Hz.

To calculate $\omega_{\mathrm{L1}}$ and $\omega_{\mathrm{L2}}$ we measure the frequencies of the mechanical modes in their thermal motion before each STIRAP sequence. The two effects described above shift the frequency of the defect mode, effectively introducing a small two-photon detuning with complicated dependence on time, which we do not take into account in our simulations. This small two-photon detuning of the order of 5Hz becomes comparable to the width of the two-photon detuning curve and starts influencing the state transfer with $\sigma\gtrsim 25$msec as the width of the two-photon detuning curve is inversely proportional to $\sigma$.


\bibliography{stirap_suppl.bib}